\documentclass[a4paper,11pt]{article}
\pdfoutput=1 

\usepackage{jinstpub} 
\usepackage[utf8]{inputenc}
\usepackage{lineno}
\usepackage{textcomp}
\usepackage{xcolor}
\usepackage{graphicx}
\usepackage{mathtools}
\usepackage{subcaption}
\usepackage{upgreek}
\usepackage{multirow}
\usepackage{arydshln}
\usepackage{doi}
\usepackage{enumitem}
\usepackage{cleveref}
\DeclareCaptionSubType*{figure}\renewcommand\thesubfigure{\thefigure\alph{subfigure}}
\usepackage{utils/general/ptdr-definitions}

\title{\boldmath A title with some math: $x=1$}

\title{Response of a CMS HGCAL silicon-pad electromagnetic calorimeter prototype to 20-300\GeV positrons}

\collaboration{CMS HGCAL collaboration}

\author[2]{B.~Acar,}
\author[13]{G.~Adamov,}
\author[37]{C.~Adloff,}
\author[26]{S.~Afanasiev,}
\author[45]{N.~Akchurin,}
\author[2]{B.~Akg\"{u}n,}
\author[4]{F.~Alam Khan,}
\author[25]{M.~Alhusseini,}
\author[5]{J.~Alison,}
\author[19]{A. ~Alpana,}
% ...

% ...
\affiliation[2]{Bo\u{g}azi\c{c}i University, \\Bebek 34342, Istanbul, Turkey}
% ...
\affiliation[4]{CERN,\\Espl. des Particules 1, 1211 Geneva 23, Switzerland}
\affiliation[5]{Carnegie Mellon University, \\ 5000 Forbes Ave, Pittsburgh 15213, PA, USA}
% ...
\affiliation[13]{Georgian Technical University, \\ 77 Kostava Str 0175, Tbilisi, Georgia}
% ...
\affiliation[19]{Indian Institute of Science Education and Research, \\ Dr. Homi Bhabha Road 411008, Pune, India}
% ...
\affiliation[25]{The University of Iowa,\\ 203 Van Allen Hall, Iowa City, 52242, Iowa, USA}
\affiliation[26]{International Intergovernmental Organization Joint Institute for Nuclear Research JINR, \\ 6 Joliot-Curie St, Dubna 141980, Moscow, Russia}
% ...
\affiliation[37]{National Central University Taipei (NCU),\\No.300, Jhongda Rd 32001, Jhongli City, Taiwan}
% ...
\affiliation[45]{Texas Tech University,\\ Lubbock 79409, TX, USA}
% ...
\emailAdd{Catherine.Adloff@cern.ch, Stathes.Paganis@cern.ch}

\abstract{
The Compact Muon Solenoid Collaboration  is designing a new high-granularity endcap calorimeter, HGCAL, to be installed later this decade.
As part of this development work, a prototype system was built,
with an electromagnetic section consisting of 14 double-sided structures, providing 28 sampling layers. Each sampling layer has an hexagonal module, where a multipad large-area silicon sensor is glued between an electronics circuit board and a metal baseplate.
The sensor pads of approximately 1.1$\cm^2$ are wire-bonded to the circuit board and are readout by  custom integrated circuits. 
The prototype was extensively tested with beams at CERN's Super Proton Synchrotron in 2018.
Based on the data collected with beams of positrons, with energies ranging from 20 to 300\GeV,
measurements of the energy resolution and linearity, the position and angular resolutions,
and the shower shapes are presented and compared to a detailed \GEANTfour simulation.}


\begin{document}
\maketitle
\flushbottom


\section{Introduction}
\label{sec:introduction}
%
%

The Compact Muon Solenoid (CMS) Collaboration will replace the existing calorimeters in the endcaps with a new high-granularity calorimeter (HGCAL) \cite{CMS_HGCAL_TDR} for the High Luminosity-LHC (HL-LHC).
It will be a sampling calorimeter with hexagonal multipad large-area silicon sensors and plastic scintillator tiles as the active media. 
The calorimeter endcaps (CE) have both electromagnetic (CE-E) and hadronic (CE-H) sections.
The absorber layers of the CE-E section are alternating plates of either lead cladded with stainless steel or copper and copper-tungsten plates; the active layers are all segmented silicon (Si) sensors. 
The CE-H section uses stainless steel as the absorber and a combination
of scintillator and silicon is used as the active material, with Si in the regions of highest radiation. The choice of this particular design was made to cope with the significantly higher radiation levels and the contribution of overlapping events (pileup) expected during the HL-LHC operation, compared to the current LHC conditions. In addition, it offers significant benefits for the reconstruction of physics objects, while providing the required tolerance to radiation damage \cite{CMS_HGCAL_TDR}. The high granularity of the detector will allow particle-flow measurements to extend from the tracker into the calorimeter, and together with the timing capability will allow for the subtraction of the energy from pileup events leading to a good energy resolution even in a high pileup environment. Merged jets can be reconstructed with higher efficiency and better energy resolution, improving the boosted object reconstruction performance. The high lateral granularity allows the tagging of narrow jets originating from the vector boson fusion production mode of the Higgs boson, as well as jets from the weak vector boson scattering  process. The high granularity also allows efficient electron and photon reconstruction in the presence of high pileup in the forward region. At the same time, the expected small constant term that typically dominates the energy resolution at high energies, will lead to an electron and photon resolution similar to the current detector. For example, the Higgs to diphoton mass resolution is not expected to be degraded by replacing the current crystal calorimeter with the HGCAL.

The production of the first HGCAL silicon sensor prototypes started in late 2015, following the original HGCAL design~\cite{CMS_HGCAL_TDR}.
Hexagonal modules were built with six-inch Si sensors. The sensors, subdived into 1.1$\cm^2$ hexagonal pads, were connected to the readout printed circuit board (PCB) with wirebonds through holes in the PCB. The signals from the pads were read by the Skiroc2 application-specific integrated circuit (ASIC) which was developed for the CALICE collaboration \cite{Skiroc2}.
Beam tests in 2016 validated the HGCAL design of the CE-E section and provided the first performance measurements and comparison with
simulation, albeit with a limited number of layers~\cite{HGCAL-2016TB}.

In October 2018, a full 28-layer electromagnetic calorimeter
was assembled with a hadronic calorimeter and placed in the Super Proton Synchrotron (SPS) H2 beamline~\cite{HGCAL-2018TB-H2} at CERN.
The CE-E prototype had a depth of 27 radiation lengths ($X_0$) and 1.4 interaction lengths ($\lambda_I$). The silicon active material, pad size, longitudinal segmentation and choice of absorber materials all followed closely the design described in \cite{CMS_HGCAL_TDR}, while the sensor size, detector length, readout electronics and data acquisition (DAQ)~\cite{HGCAL-2018TB-H1} all differed from the original design. An improved version of the front-end ASIC, the Skiroc2-CMS~\cite{Skiroc2}, with a large dynamic range and timing information was used.
Measurements were made with asynchronous beams of muons, positrons and charged pions, at a low enough rate that there were few overlapping particles.
One of the main aims of the 2018 tests was to study in detail the performance of the CE-E prototype with beams of positrons with energies ranging from 20\GeV to 300\GeV.

In this paper, the results of the tests with positrons are presented. The linearity and resolution of the measured energy and the reconstructed positron energy, the position and angular resolutions and the longitudinal and lateral shower shapes are discussed. 
The details of the experimental setup are described in Section~\ref{sec:setup}, the data collected and the Monte Carlo simulation (MC) are presented in Section~\ref{sec:datasimu}, and the analysis framework is outlined 
in Section~\ref{sec:frameworkpresel}.
The main results obtained are presented in 
Sections~\ref{sec:energyrl}-\ref{sec:showershape} and 
in Section~\ref{sec:conclusion} a summary of the results is given.

\section{Beam Tests Experimental Setup}
\label{sec:setup}
%
%
A detailed understanding of the beamline and of the prototype, including its trigger system are mandatory for the proper modeling in the simulation and for the interpretation of the results. In this section, we present their relevant characteristics.
\subsection{CERN Beamline}
\label{beamline}
The H2 beamline \cite{h2-beamline} is located in the CERN-SPS North Area. 
Secondary beams of hadrons, electrons and muons with up to a maximum momentum of 300\GeVc and 380\GeVc for hadrons, with both charges available, are transported in the experimental areas with variable purities (10--99\%) and rates.
These secondary beams are produced by the interaction of the primary proton beam impinging on a 500\mm thick beryllium 
primary target. The H2 beamline transports the
produced particles over a length of approximately 590\meter, with the
last 180\meter being inside the EHN1 experimental hall. 
The H2 beamline is a magnetic spectrometer consisting of dipole and quadrupole magnets and collimators.
The beamline selects the secondary particles produced at the target within a relative momentum acceptance of 0.2--2\%, depending on the collimator settings.
The currents of the first and second sets of dipole magnets define the beam momentum of the particles emerging 
from the last dipole magnet of the spectrometer located 240\meter upstream of the HGCAL prototype.
The accuracy in the setting of the current in the dipole magnets corresponds to an uncertainty in the momentum of $\pm 1$\GeV. The final beam is achromatic
to first order, with a fixed momentum spread defined by the collimators. Second-order chromatic aberrations introduce a negligible correlation between each particle momentum and its transverse position.
For hadrons, there is no other contribution to the spread in momentum. However, for positrons, synchrotron radiation (SR) losses in the dipole magnets, particularly important for energies above 100\GeV, induce an additional beam momentum spread, leading to a final beam momentum that is systematically lower than the nominal one.
Calculated from a simulation of the full beamline, described in Section~\ref{sec:datasimu}, the positron beam momentum and its spread as a function of the nominal momentum are given in Table~\ref{tab:BeamEnergiesFullBL}. The spread observed at low energy is due to bremsstrahlung losses and to losses from interactions of the positrons with material in the straight section of the beamline upstream of the calorimeter.
\begin{table}[!ht]
\begin{center}
  \caption{
  H2 beamline positron momentum and its spread as a function of the nominal momentum, calculated from a simulation of the full beamline described in Section~\ref{sec:datasimu}. 
  The final beam momentum and its spread, were taken from the mean and the standard deviation of a Gaussian fit to the sum of particle momenta at the entrance of the CE-E prototype. The fit was performed iteratively within two standard deviations from the mean, i.e. where the distribution is Gaussian. The momentum resolution was defined as the ratio of the standard deviation to the mean.}
  \begin{tabular}{|c|c|c|c|}
  \hline         
    Nominal &Final    & Final & Final  \\
    Momentum&Momentum & Momentum Spread & Momentum Resolution\\
    $\text{[}$\GeVc] & [\GeVc] & [\GeVc] & \% \\
    \hline
       20   &   20.00   & 0.06 & 0.3\\
       30   &   30.00   & 0.08 & 0.3\\
       50   &   49.97   & 0.12 & 0.3\\
       80   &   79.91   & 0.19 & 0.2\\
      100   &   99.81   & 0.22 & 0.2\\
      120   &   119.64  & 0.28 & 0.2\\
      150   &   149.16  & 0.35 & 0.2\\
      200   &   197.40  & 0.47 & 0.2\\
      250   &   243.84  & 0.60 & 0.2\\
      300   &   287.65  & 0.79 & 0.3\\ \hline
  \end{tabular}
  \label{tab:BeamEnergiesFullBL}
\end{center}
\end{table}

\begin{figure*}[!ht]
  \begin{center}
    \includegraphics[width=1.0\textwidth]{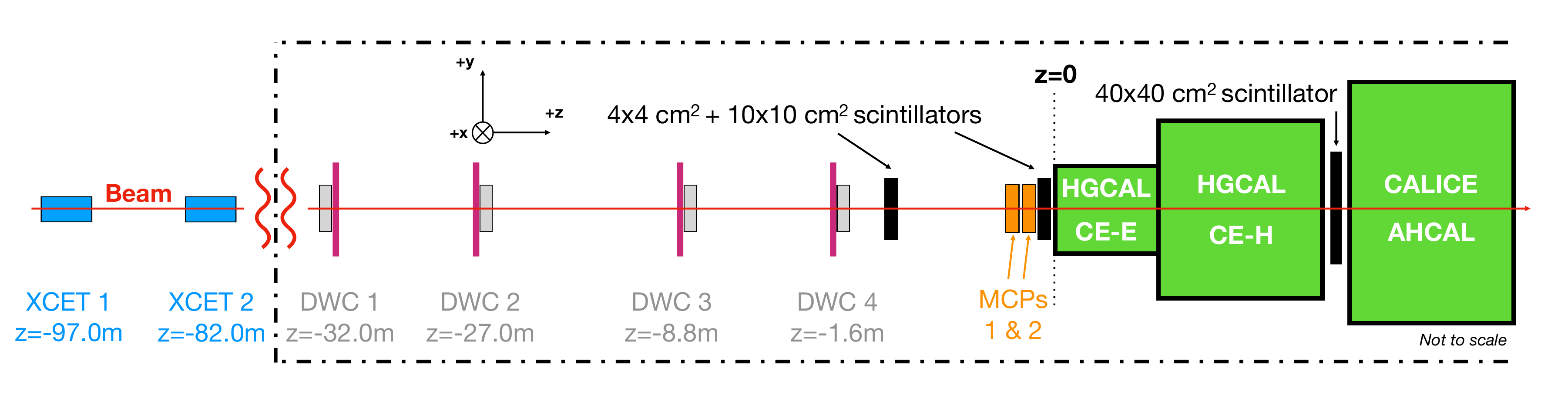}
  \end{center}
\caption{H2 beamline and HGCAL-prototype setup.}
\label{fig02:setup}
\end{figure*}

The setup of the H2 beamline and HGCAL-prototype are shown in Fig.~\ref{fig02:setup}. 
From the last spectrometer dipole magnet to the face of the calorimeter, there are air gaps, beam windows and beam counters, including delay wire chambers (DWC) and Cherenkov (XCET) detectors, totalling approximately 0.5~$X_0$ of material.

\subsection{HGCAL Prototype}
\label{Prototype}
The prototype of the HGCAL comprised a CE-E and a CE-H section. 
The CE-E section consisted of sampling layers of hexagonal modules with hexagonal Si pads ($\approx$ 1.1$\cm^2$ per pad) interleaved with alternating
copper and copper-tungsten absorbers or lead and stainless-steel absorbers. 
For the 2018 beam test, 28 hexagonal modules were assembled as a glued stack of a copper-tungsten baseplate (except for modules 21 to 24), a Kapton foil, a silicon sensor and a readout PCB, the `Hexaboard'. 

The modules followed the same basic design as the prototypes tested in 2016, with improved grounding to reduce the electronic noise and the new Skiroc2-CMS front-end chip \cite{Skiroc2-CMS} that was specifically designed to meet the HGCAL requirements. It has a large dynamic range for energy measurements, thanks to a dual-gain amplifier, and a timing chain that provides a time-over-threshold measurement (ToT) to cope with very high energy deposits when the low-gain chain is saturated. 
For the low- and high-gain chains, signals collected from each Si pad are amplified, shaped, sampled with a 25~ns frequency
and stored in a 
13-bin switched capacitor array rolling analogue memory (SCA).
The ASIC also provides a time-of-arrival measurement (ToA) to study the feasibility of precise time measurements to contribute to pileup rejection
\cite{HGCAL-2018TB-H3}.
These changes are aimed to approach the HGCAL final design.

\begin{figure}[!t]
  \begin{center}
    \includegraphics[width=1.07\textwidth, angle=-90]{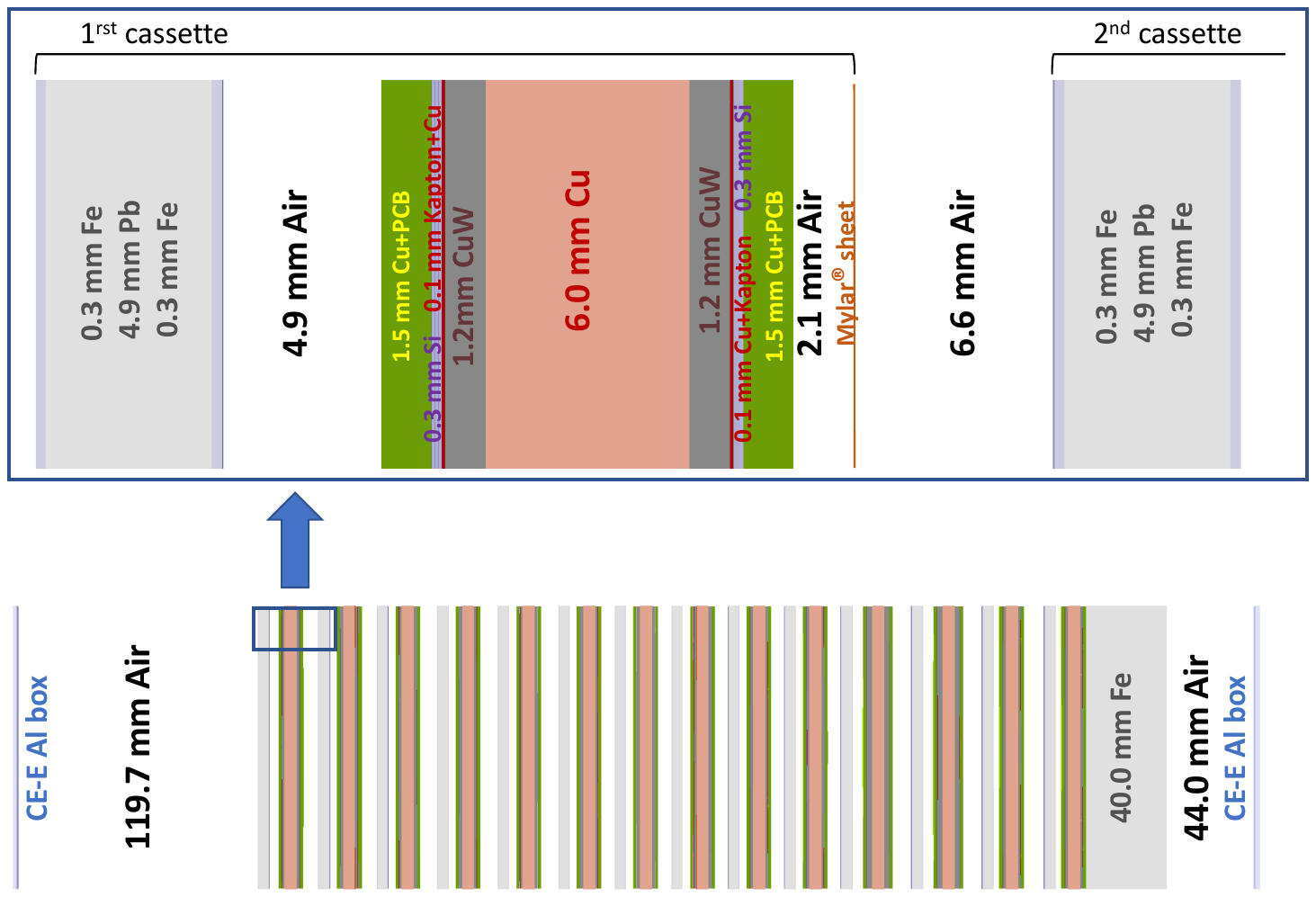}
  \end{center}
\caption{ 
Layout of the CE-E prototype. This prototype was built with 14 cassettes, each with two hexagonal modules. In the beam test, the beam entered the prototype through the first cassette (top in the figure). 
In an hexagonal module, the Si sensor was between the copper-tungsten plate (dark grey, CuW) and the Hexaboard (green, Cu+PCB). 
A cassette started with a lead plus stainless-steel absorber (light blue and light grey, Fe+Pb+Fe) and ended at the Hexaboard of the second module. The cassette end was enclosed by a Mylar sheet, held by an aluminium frame (not represented here).
The zoomed-in area shows the first cassette. The lead plus stainless-steel absorber of the second cassette is also shown including the air gap between the two cassettes. Through the calorimeter, this air gap varied between 4.6mm and 6.9mm. The two drawings are scaled proportionally.}
\label{fig02:CEE-layout}
\end{figure}
The flexibility of the prototype mechanical assembly allowed for different configurations of the detector to be studied.
The results presented in this work are based on the data taken with a configuration where the CE-E prototype had 14 double-sided structures (referred to as cassettes) or 28 sampling layers, as shown in Fig.~\ref{fig02:CEE-layout}. 
All the silicon sensors were made with 300~$\upmu$m thick wafers, while the last two were 200~$\upmu$m thick.
Table~\ref{tab:dEdxWeights} gives the corresponding peak value of the simulated energy loss in the sensor by a 150\GeV muon with normal incidence, $\Delta E_{i}^{\text{Si}}$,
determined by a fit to the energy distribution with a Landau convoluted with a Gaussian function. 
The depth, $z_i$, in radiation lengths, before the $i^{\text{th}}$ sensor  is also given, as well as the thickness, $\Delta z_i$, in radiation lengths between two sensors and the corresponding mean minimum ionisation energy loss of a muon, $\Delta E_{i}^{\text{Abs}}$, from \cite{PDG}.
The amount of material between sensors follows an odd/even pattern according to the design of the CE-E prototype. Exceptions are the first absorber layer that included the aluminium container of the prototype, and layers 22 and 24 which had more material.
The modules 21 to 24 were assembled on a 1.2~mm copper baseplate instead of the a CuW plate, and a 1.2~mm Cu/W plate was added to increase the mean energy loss for the absorbers 22 and 24.
Further details on the CE-E and CE-H sections can be found in \cite{HGCAL-2018TB-H2}.
\begin{table}[!t]
  \begin{center}
    \caption{The characteristics of the 28 layers of the 14-casettes configuration of the CE-E prototype. 
    The second column gives the peak value of the simulated energy loss in the Si sensor by a 150\GeV muon. The third and fourth columns give the calorimeter depth before each sensor and the thickness between two sensors in units of radiation length.
    The fifth column gives the mean minimum ionisation energy loss of a muon in the absorber layers computed with \cite{PDG}.
    The last column gives the average of this energy loss in the absorber before and after the sensor,
    except for the last layer where $\Delta E_{28}^{\text{Abs}}/2$ is given. 
    The sum of the second and last column is the  weight for each layer given by Eq.~\ref{eq:Abs_Energy_Rec_Detailed}.
    }
\begin{tabular}{ |c c|c|c|c|c|c|}
  \hline
\multicolumn{2}{|c|}{Cassette}   & Energy Loss          & Depth& Absorber Thickness& Energy Loss & Average   \\
\multicolumn{2}{|c|}{/ Layer}  & $\Delta E_{i}^{\text{Si}}$ & $z_i$& $\Delta z_i$ &  $\Delta E_{i}^{\text{Abs}}$ & $\big(\Delta E_{i}^{\text{Abs}} + \Delta E_{i+1}^{\text{Abs}}\big)/2$   \\
\multicolumn{2}{|c|}{$h/i$}     & [MeV]             & [$X_0$] & [$X_0$] &  [MeV]                                                  & [MeV]               \\
\hline
\multirow{2}{*}{1}& 1   &  0.085  &   1.00 & 1.00  &  10.2 &  11.29 \\
 &2   &  0.085  &   1.98 & 0.98  &  12.3 &  9.85  \\ 
\hdashline
\multirow{2}{*}{2}& 3   &  0.085  &   2.92 & 0.94  &  7.4  &  9.85  \\ 
&4   &  0.085  &   3,90 & 0.98  &  12.3 &  9.85  \\ 
\hdashline
\vdots & \vdots & \vdots & \vdots & \vdots & \vdots & \vdots\\
\hdashline
\multirow{2}{*}{10}& 
19  &  0.085  &  18.23 & 0.94  &  7.4  &  9.85  \\ 
&20  &  0.085  &  19.21 & 0.98  &  12.3 &  9.85  \\
\hdashline
\multirow{2}{*}{11}& 
21  &  0.085  &  20.15 & 0.94  &  7.4 &  11.36 \\ 
&22  &  0.085  &  21.30 & 1.15  &  15.4 &  11.36 \\ 
\hdashline
\multirow{2}{*}{12}& 
23  &  0.085  &  22.23 & 0.94  &  7.4  &  11.36 \\ 
&24  &  0.085  &  23.38 & 1.15  &  15.4 &  11.36 \\ \hdashline
\multirow{2}{*}{13}& 
25  &  0.085  &  24.31 & 0.94  &  7.4  &  9.85  \\ 
&26  &  0.085  &  25.29  & 0.98  &  12.3 &  9.85   \\\hdashline 
\multirow{2}{*}{14}& 27  &  0.057  &  26.23 & 0.94  &  7.4  &  9.85   \\ 
&28  &  0.057  &  27.21 & 0.98  &  12.3 &  6.17   \\ 
\hline
\end{tabular}
\label{tab:dEdxWeights}
\end{center}
\end{table}

The acquisition of the positron data was triggered by the coincidence of two scintillator counters upstream of the calorimeter and the veto of a scintillator counter downstream of the CE-H section. An event corresponds to the data recorded after a coincidence of the trigger counters. 
A complete description of the data acquisition system can be found in 
\cite{HGCAL-2018TB-H1}.

\section{Data and Simulation Samples}
\label{sec:datasimu}
The detector was exposed to beams of muons, positrons and pions. In this publication, only the data collected with the positron beam are discussed, with nominal beam energies ranging from 20\GeV to 300\GeV.
Around 100k events per positron beam energy setting were recorded. 

An event selection, defined in Section~\ref{PreSelection},
was used to obtain event samples with a high positron purity, with minimal energy loss upstream of the prototype, and with full lateral containment of the positron shower within the prototype.
Between 30\% and 75\% of the recorded events were selected depending on the beam energy and beam settings. The main event loss arose from a geometric cut on the impact position of the impinging track.

The three detector setup configurations were simulated with the \GEANTfour toolkit version 10.4.3, \cite{Geant4},  
using the FTFP\_BERT\_EMN physics list. This simulation includes details of the CE-E prototype, from the composition and thickness of each material to the air gap between the cassettes.
The input of the HGCAL prototype simulation (beam gun) was a multiple particle generation  constructed with the output of the particle tracking simulation program {\textsc{G4BeamLine}}\xspace based on \GEANTfour \cite{G4beamline}.
The physics list used for the beamline simulation was FTFP\_BERT\_EMZ.
The beam content at the exit of the target was set to 90\% of positrons and 10\% of protons. At the calorimeter, the proton contamination of the beam is negligible for energies larger than 150\GeV. The estimation of the initial proton contamination has a minor influence on the final results due to the event selection described in Section~\ref{PreSelection}. 
All the relevant H2 beamline elements (passive and active) were included from the exit of the production target to the HGCAL prototype, including quadrupole/dipole magnets, bending magnets, collimators, beam windows, beam pipes, scintillator counters, air sections (40\% relative humidity), materials of fixed experiments NA61/SHINE and DWCs, which were all present in the beamline.
When compared with earlier, less detailed simulations of the last 30\meter of the H2 beamline only, this new simulation framework provided a better  description of the beam characteristics at the entrance of the prototype, agreeing well with the observed characteristics despite the assumption on the initial particle production. 
For example, 
the simulation generated a beam position spread at the entrance of the HCGAL prototype that was in close agreement with the data.

The offsets in the horizontal ($x$) and vertical ($y$) directions and the angular displacement measured with the CE-E prototype, were all included in the beam gun used in the the detector simulation.
The simulation of the calorimeter response included the requirement that the trigger conditions are met. The number of events simulated was 100k for each beam energy, similar to the number of events collected.

In the detector simulation, the effect of the intrinsic electronics noise was included at the level of the simulated energy deposited in a Si pad, and non-responsive pads or defective electronic channels were masked. Pad non-uniformity in response,
pad-to-pad crosstalk, digitisation, and other electronics effects were, however, not included in the simulation. 

To study the effect of the crosstalk on the transverse shower shape measurements, a specific simulation was used, that included the crosstalk between every pad and its nearest neighbours, without any charge dependence. 
The amount of crosstalk was estimated using a charge injector to deliver pulses directly to the inputs of the ASIC where they were bonded to the sensor pads. It was found that every pad capacitively induces a total charge in the surrounding 6-pad ring, of approximately 5\% of the injected charge. This percentage increases very slowly with the amplitude of the pulse and varies very little from pad to pad. The few unbonded channels showed no significant crosstalk.

The DWCs were not included as sensitive materials in the full beamline simulation.
Instead, 
in the simulation, the impact position at each DWC was extrapolated backwards in the direction of the most energetic charged particle at the front face of the CE-E.
The impact positions were smeared in the $x$ and $y$ directions with a Gaussian function with the estimated DWC intrinsic resolution of 500~$\upmu$m, to obtain the simulated DWC measurements \cite{T_Quast_Thesis}. This results in a 430~$\upmu$m precision on the impact position of the reconstructed DWC track at the level of the CE-E prototype.

\section{Analysis Framework}
\label{sec:frameworkpresel}
%
%
In this section, we provide a brief description of the reconstruction framework and the criteria used to select data for the analysis.

\subsection{Reconstruction Framework}
\label{Framework}
The data from Skiroc2-CMS ASICs consisted of the digitized readouts of the 13 SCA elements of the high-gain (HG) and low-gain (LG) channels, and of the ToT and ToA readouts.
The pad locations in right-handed Cartesian coordinates, 
$\left(x_i,y_i,z_i\right)$, were retrieved from the corresponding Hexaboard number, chip number and channel number. 
The location of a sensor in the longitudinal direction ($z$) is known to a precision of 0.5\mm, and the location of a pad within a Si sensor is known to a precision of 0.1\mm in $x$ and $y$.

\begin{figure*}[!ht]
  \begin{center}
  \begin{minipage}[r]{0.45\textwidth}
      \vspace{0pt}
      \includegraphics[width=0.95\textwidth]{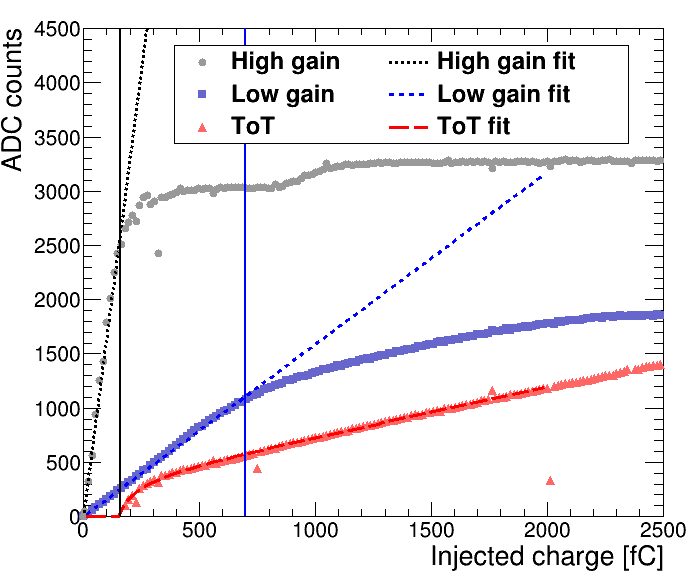}
      \vspace{-20pt}
      \end{minipage}
  \end{center}
  \caption{High-gain and low-gain amplitudes and ToT output as a function of the input charge.
  The vertical solid lines mark the maximum input signals where the high-gain and low-gain shaper outputs were linear.}
  \label{fig:ReadoutGain}
\end{figure*}
The reconstruction steps to obtain in data the energy deposited in a Si pad are described in detail in~\cite{HGCAL-2018TB-H2}, they are summarized below here:
\begin{itemize}
\item
Pedestals were determined for each channel and were subtracted from each readout of the SCA.
Then, the noise common to every channel in each hexagonal module, 
was estimated for each time sample and subtracted. This procedure was followed because there was a high correlation of the noise in the four ASICs on a module, with a high frequency component in the resulting common-mode noise ~\cite{HGCAL-2018TB-H2}.
\item
The waveforms were reconstructed from the time-ordered corrected SCA values.
For a pad, a hit was defined when 
a pulse was identified in the HG reconstructed waveform.
By fitting to the waveforms, the HG and LG amplitudes, $A^{\text{HG}}$ and $A^{\text{LG}}$, were determined for each hit.
The resulting signal-to--noise ratio (after the common-mode noise subtraction) for 200 GeV muons is, for most of the layers, larger than 6 for the HG and 3 for the LG chains~\cite{HGCAL-2018TB-H2}.
For the ToT, the offset-subtracted readout values gave directly an estimate of the signal amplitude $A^{\text{ToT}}$. 
\item
The Si pad response, $ A_{\text{HG}}'$, was computed using the best value determined from the linear regions of HG, LG and ToT channels:
\begin{equation}
    A_{\text{HG}}' = \left\{ \begin{array}{ll}
    A^{\text{HG}}                  & \text{, if } A^{\text{HG}}<\text{HG}_{\text{sat}}\\
    A^{\text{LG}}  \cdot m_{\text{HG}/\text{LG}} & \text{, if } A^{\text{HG}}>\text{HG}_{\text{sat}} \text{  and } A^{\text{LG}}<\text{LG}_{\text{sat}}\\  
    A^{\text{ToT}} \cdot m_{\text{LG}/\text{ToT}} \cdot m_{\text{HG}/\text{LG}} & \text{, otherwise} 
    \end{array} \right.
  \label{eq:gains}
\end{equation}
where, $m_{\text{LG}/\text{ToT}}$ and $m_{\text{HG}/\text{LG}}$ are scale factors from \text{LG} to ToT and from \text{HG} to \text{LG}, respectively. $\text{HG}_{\text{sat}}$ is the threshold above which the $A^{\text{LG}}$
was used, and $\text{LG}_{\text{sat}}$ is the threshold above which the $A^{\text{ToT}}$ was used, as illustrated in Fig.~\ref{fig:ReadoutGain}.
Using the positron data, these parameters were determined for as many readout channels as possible. 
For channels away from the beam axis, with low hit occupancy, the average value of the parameters for the corresponding chip or Hexaboard was used.
\item
To insure the uniformity of the measured energy for all pads, an intercalibration factor, $M_{\text{HG}/\text{MIP}}$, was determined with 200\GeV muons with normal incidence, which we label as MIP\footnote{A MIP in this case is a 200\GeV muon which has approximately 7\% more energy loss than the mininum ionising particle.}, and used to correct for response variations.
The $M_{\text{HG}/\text{MIP}}$ value, referred to as MIP value, was the peak value determined for each pad by the fit to the HG amplitude distribution with a Landau convoluted with a Gaussian function (in this case the peak and the most probable values differs). The intercalibrated pad energy in MIP units was given by:
\begin{equation*}
        E_{\text{pad}}^{\text{Si}}[\text{MIP}]~=~\dfrac{A_{\text{HG}}'}{M_{\text{HG}/\text{MIP}}}.
\end{equation*}
\end{itemize}

In the simulation, to allow for comparison with data, the true energy deposited in a Si-pad, smeared by the electronic intrinsic noise, was expressed in terms of energy deposited by 150\GeV muons.
The value used for the conversion, $\Delta E_{i}^{\text{Si}}$, took account of the two different Si thicknesses of the sensors and is listed in Table~\ref{tab:dEdxWeights}.

For both data and simulation, the upstream trajectory of the beam particle was reconstructed from the DWC measurements.
For data, the alignment corrections for translation offsets in the $x$ and $y$ directions were derived by comparing the reconstructed position in the first layer of the calorimeter with the position extrapolated from DWC track.

\subsection{Hit and Event Selection}
\label{PreSelection}
\begin{figure}[!ht]
\centering
\begin{subfigure}{.45\textwidth}
  \includegraphics[width=0.95\textwidth]{
  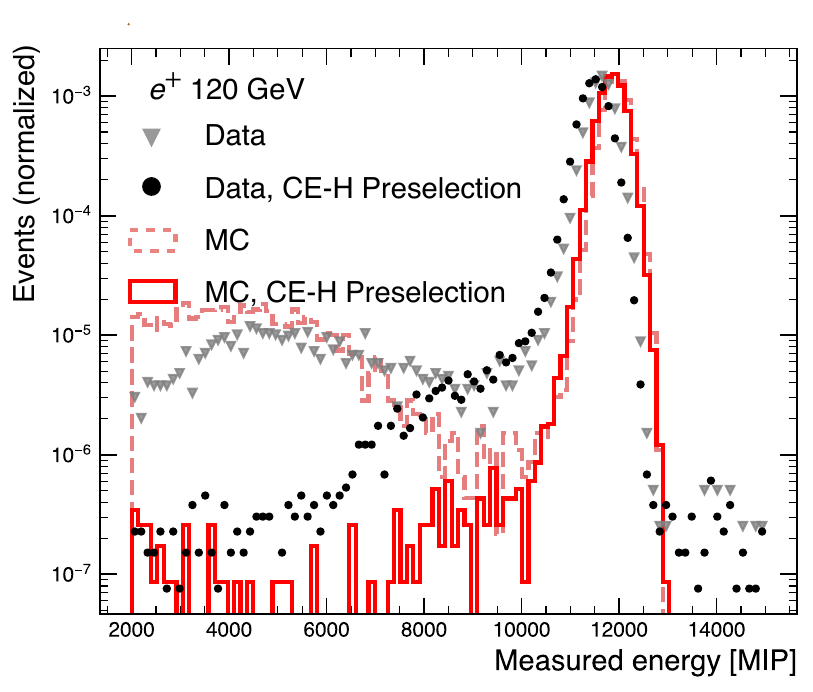}
  \caption{\label{fig:pionsRejectionDWCCutA} Comparison of the measured energy for selected hits in the CE-E prototype before and after the cuts described in \cref{one} using the \mbox{CE-H} prototype information. Hadron contamination in the positron beam is suppressed. The nominal positron energy is 120\GeV.} 
\end{subfigure}
\qquad
\begin{subfigure}{.45\textwidth}
  \includegraphics[width=0.95\textwidth]
   {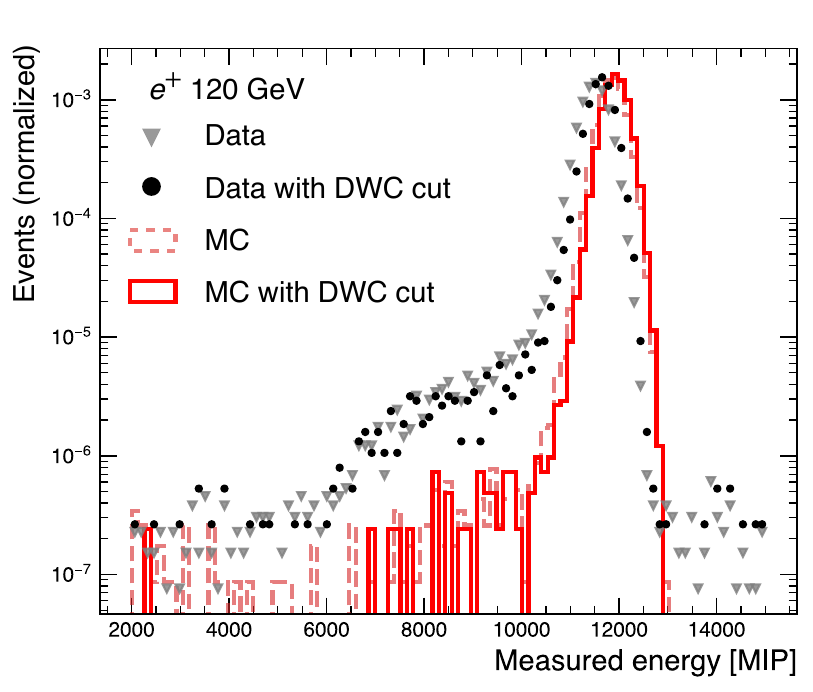}
  \caption{
  \label{fig:pionsRejectionDWCCutB} For selected hits and after the \mbox{CE-H} selection, comparison of the measured \mbox{energy} before and after the DWC acceptance cut des\-cribed in \cref{two}. The nominal positron energy is 120\GeV. \newline } 
\end{subfigure}
\end{figure}

Selections were applied equally in both data and simulation to hits and events.
The criteria to select pads with a signal were:
\begin{enumerate}[label=\alph*)]
\item
  The pad energy was required to be above 0.5~MIPs. This threshold was chosen to be well above the typical noise level of ~1/7 MIP measured in the high-gain chains of the CE-E section ~\cite{HGCAL-2018TB-H2}.
\item
  Response from abnormally noisy channels, or from the channels of one defective ASIC in the first layer were excluded. This selection affected 1.6\% of channels.
\end{enumerate}
The following criteria were used to select the events:
\begin{enumerate}[label=\roman*)]
\item
One and only one track was found in the DWCs. This requirement reduced the contribution of the events where the shower started upstream of the detector. 
\item \label{one}
Events with signals in more than 50 pads in the hadronic section 
were rejected. Additionally events where less than 95\% of the total measured energy was in the CE-E section were suppressed.
This selection removed events from pion and proton contamination in the beam ~\cite{T_Quast_Thesis}.
The effect of this requirement on the measured energy in the CE-E can be seen in Figure~\ref{fig:pionsRejectionDWCCutA}, for positrons with a nominal energy of 120~\GeV, where the hadron contamination was particularly significant (around 9\% at the entrance of CE-E, determined from the beamline simulation).
\item \label{two}
  Events with a DWC extrapolated track at the entrance of the CE-E prototype falling outside of a central $2\times 2$~cm$^2$ acceptance window were rejected. This selection limits lateral energy losses in cases where the impact is far from the prototype centre, as well as the effect of different beam profile/impact in data and MC.
  The acceptance window position was chosen to avoid events for which the centre of the shower was in close proximity to a specific pad with a defective amplifier, located in one of the layers with the highest energy contribution. 
  The effect of this acceptance cut on the measured energy can be seen in Fig.~\ref{fig:pionsRejectionDWCCutB}, for positrons with a nominal energy of 120~\GeV, showing an increase of the mean of the measured energy.
\end{enumerate}

The small shift in the peak value in the energy distribution between data and simulation, observed in Fig.~\ref{fig:pionsRejectionDWCCutA} and Fig.~\ref{fig:pionsRejectionDWCCutB}, is discussed in Section~\ref{sec:energyrl}. 

\section{Energy Linearity and Resolution}
\label{sec:energyrl}
%
An important design criterion of HGCAL is to have approximately the same energy resolution for high energy electrons and positrons as the current detector.
The results presented here were obtained from a prototype with the same number of layers as the HGCAL design described in \cite{CMS_HGCAL_TDR}. %

\begin{figure*}[!ht]
  \begin{center}
    \includegraphics[width=0.45\textwidth]{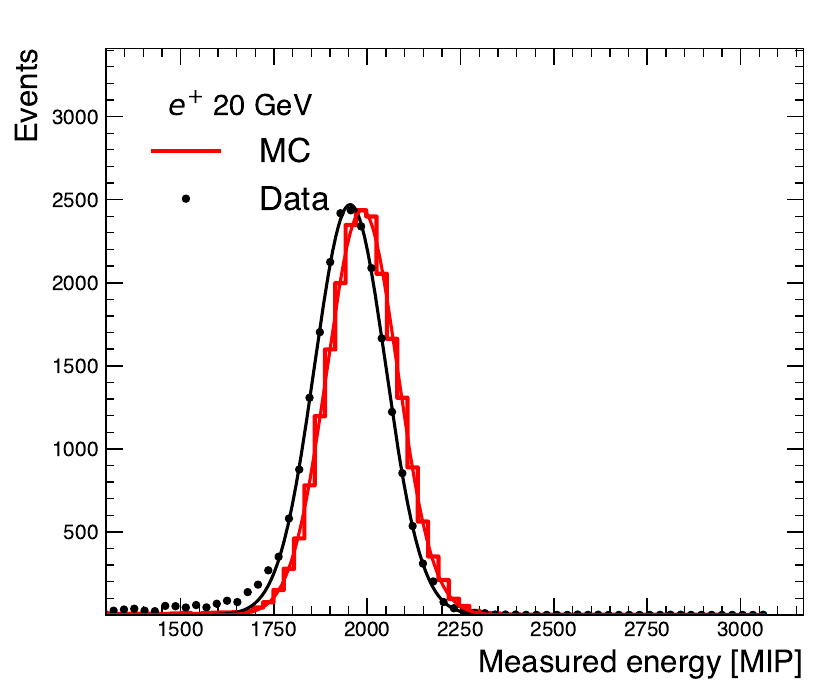}\hspace{-0.35cm}
    \llap{\raisebox{2.63cm}{
      \includegraphics[height=2.6cm]{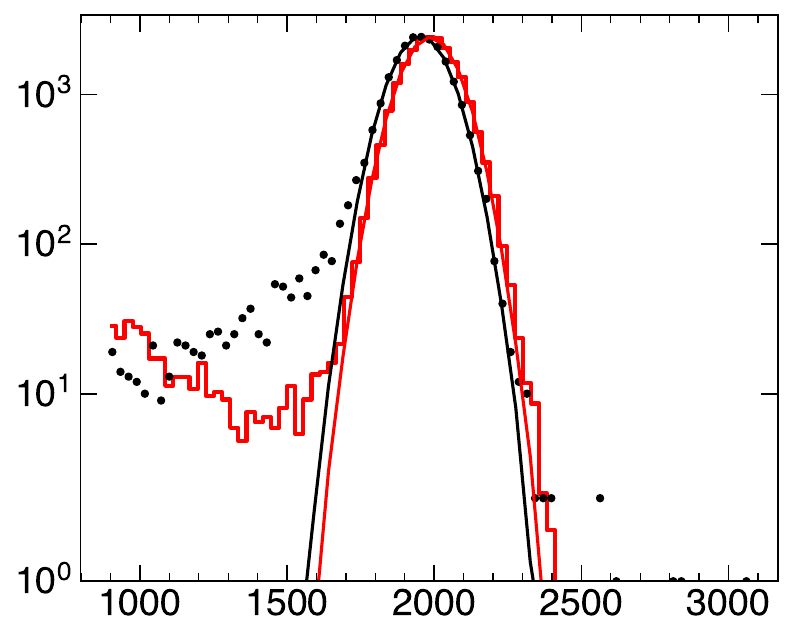}%
    }}
    \includegraphics[width=0.45\textwidth]{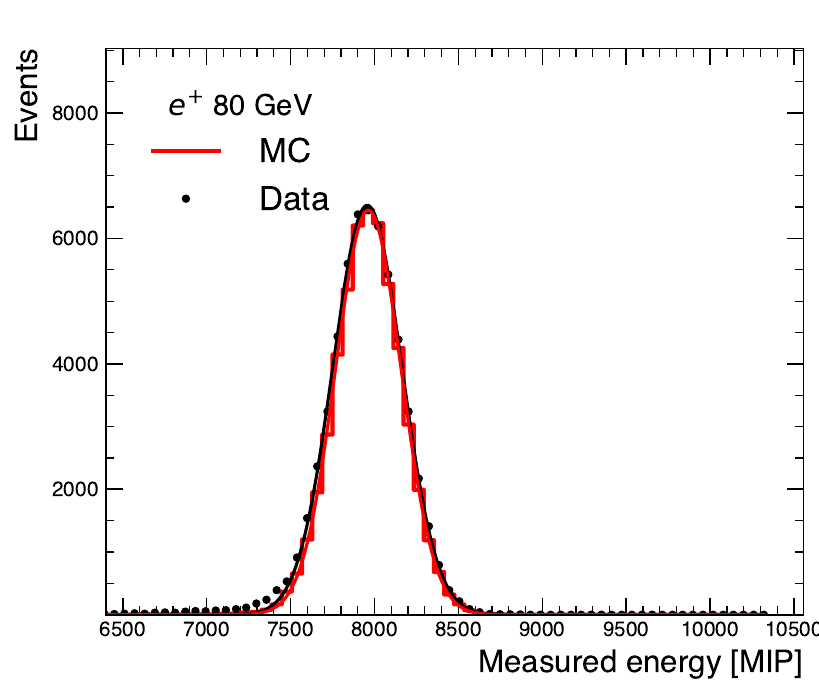}\hspace{-0.35cm}
    \llap{\raisebox{2.63cm}{
      \includegraphics[height=2.6cm]{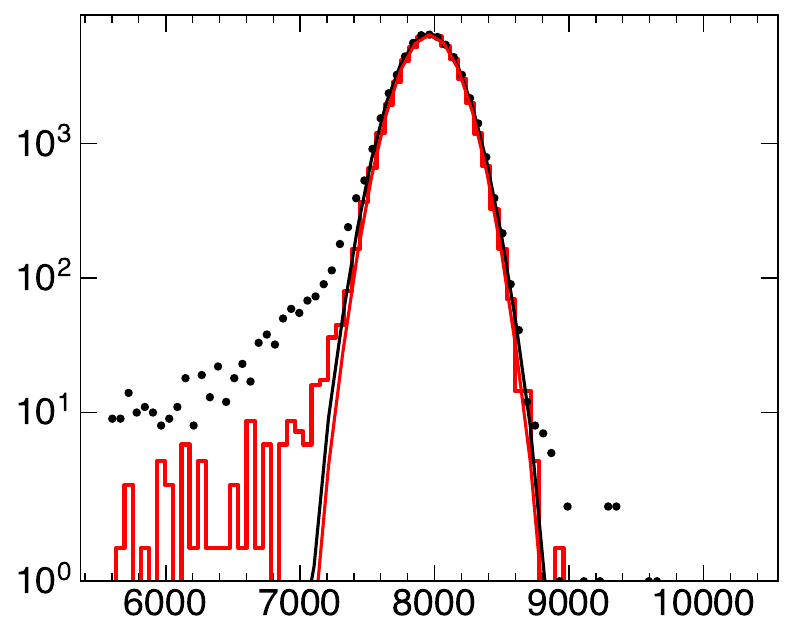}%
    }}
    \includegraphics[width=0.45\textwidth]{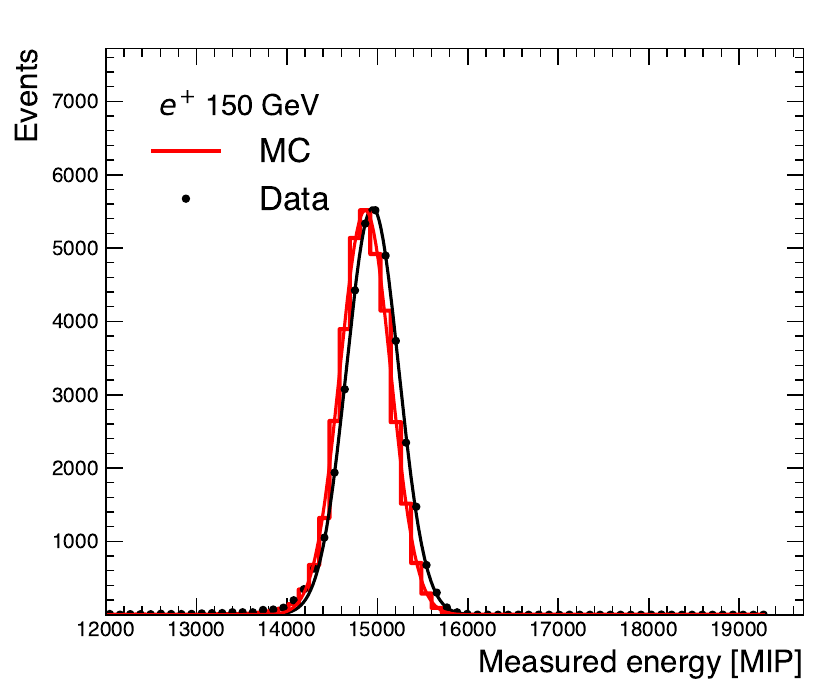}\hspace{-0.35cm}
    \llap{\raisebox{2.63cm}{
      \includegraphics[height=2.6cm]{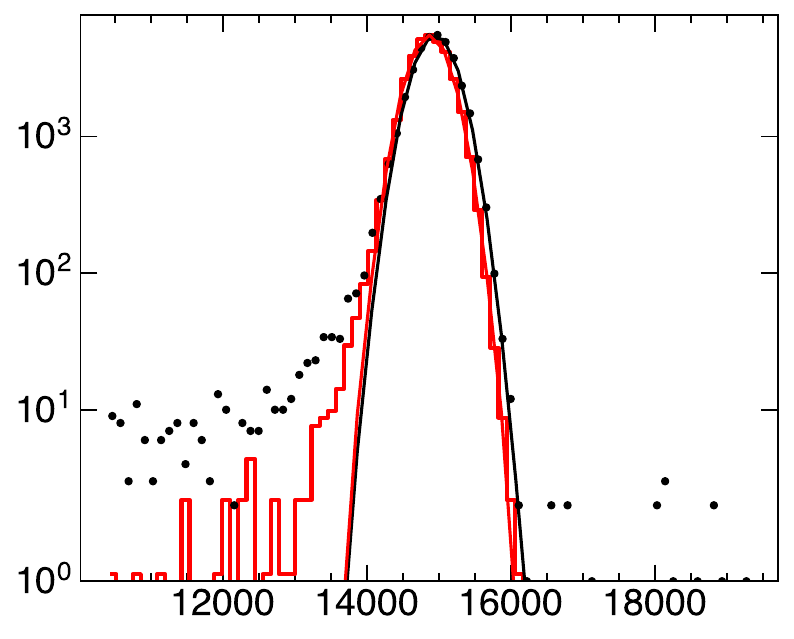}%
    }}
    \includegraphics[width=0.45\textwidth]{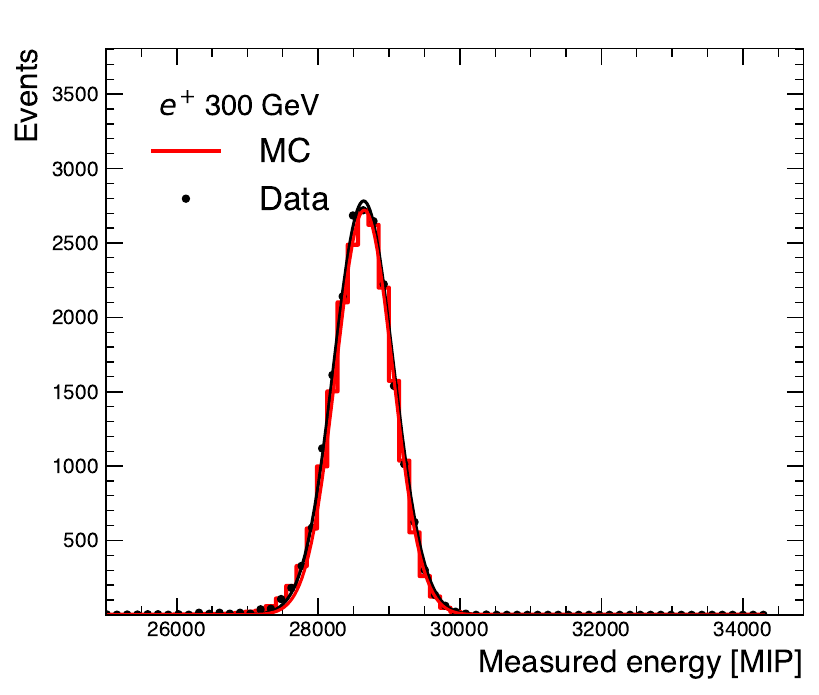}\hspace{-0.35cm}
    \llap{\raisebox{2.63cm}{
      \includegraphics[height=2.6cm]{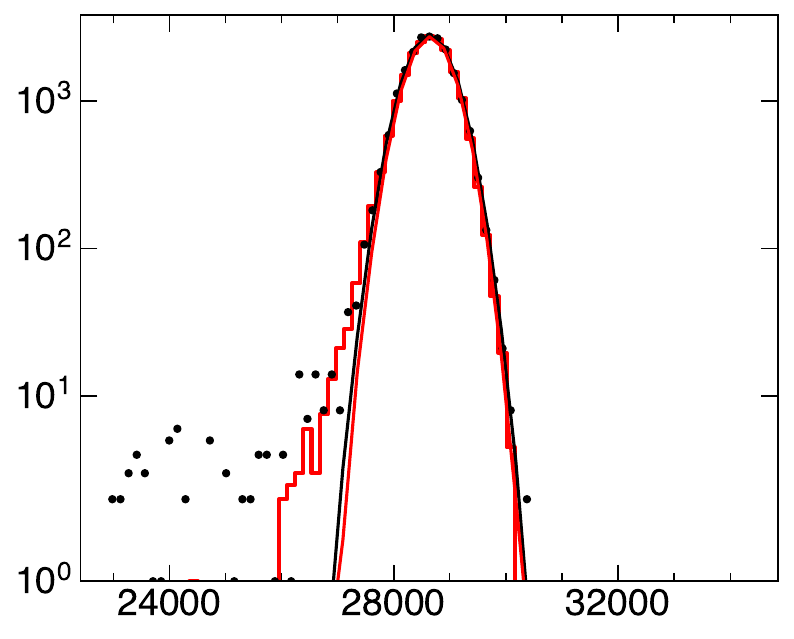}%
    }}
  \end{center}
  \caption{Measured energy distributions for data and simulation (normalized to the number of events in data) for nominal positron beam energies ranging from 20 to 300~\GeV.
  In this figure, a scale factor of 1.035 has been applied to the data.
  }
  \label{fig:dataMCvis}
\end{figure*}

Following the procedure described in Section~\ref{Framework} and the selection defined in Section~\ref{PreSelection}, the unclustered measured energy in MIP units was taken to be the sum of the energy deposited in the silicon pads of the CE-E.
The measured energy distributions in data and simulation are shown in Fig.~\ref{fig:dataMCvis} for four positron energies, where a scale factor of 1.035 has been applied to the data to match the simulation.
For each positron energy, the data and MC distributions were fit iteratively until convergence with a Gaussian function, with mean value, $\mu$, and standard deviation, $\sigma$, within the range $[\mu - 1.0\sigma, \; \mu + 2.5\sigma]$. 
The asymmetric interval was chosen to avoid biasing the mean and the standard deviation by a low energy tail in the distributions that is due to 
losses in the straight section of the beamline upstream of the calorimeter. 
The $\mu$ and $\sigma$ values from the fit were taken as the mean energy response,  $\left\langle E\right\rangle$, and resolution, $\sigma_{E}$, of the detector at a given energy, since the beam energy spread is negligible.
As discussed in Section~\ref{sec:setup}, the estimation of the final beam energies at the entrance of the calorimeter took into account systematic SR losses which is up to 4\% of the beam energy at 300~\GeV. 
The systematic 
uncertainties
in the
beam energy from uncertainties in the magnetic field of the dipole magnets and from the uncertainties in the beamline material budget were neglected.

\begin{figure*}[!ht]
\centering
\begin{subfigure}{.44\textwidth}
  \includegraphics[width=0.93\textwidth]{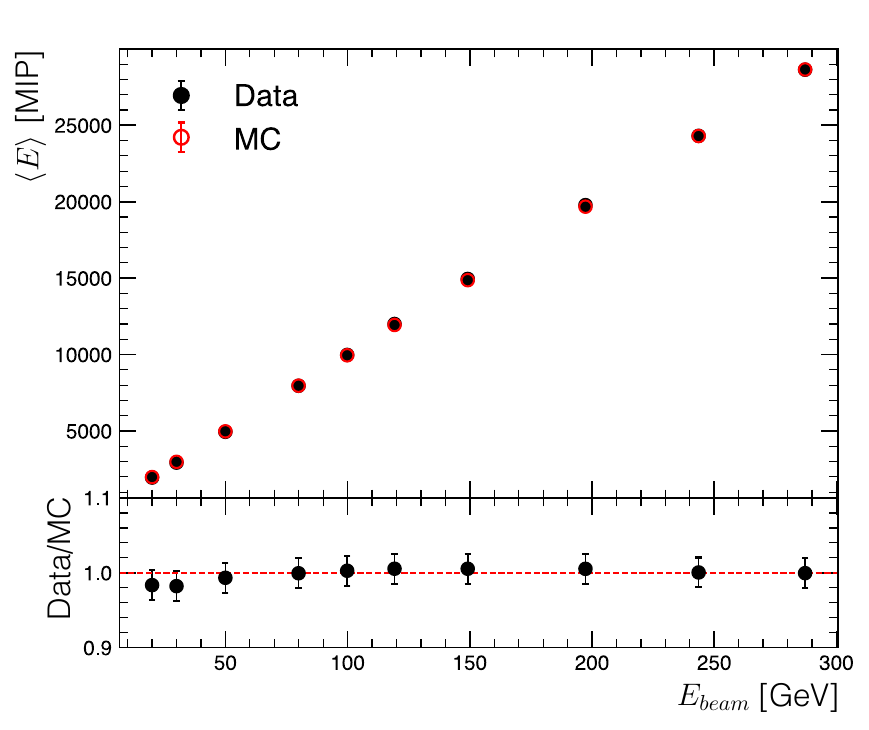}
  \caption{\label{fig:ScaleFactor} Mean measured energy as a function of the beam energy for data and simulation after applying a scale factor of 1.035 to the data. The ratio of the data to the MC mean measured energy is displayed in the lower panel (error bars are evaluated by propagating the errors on the mean measured energies). \ \newline} 
\end{subfigure}
\qquad
\begin{subfigure}{.44\textwidth}
  \includegraphics[width=0.99\textwidth]{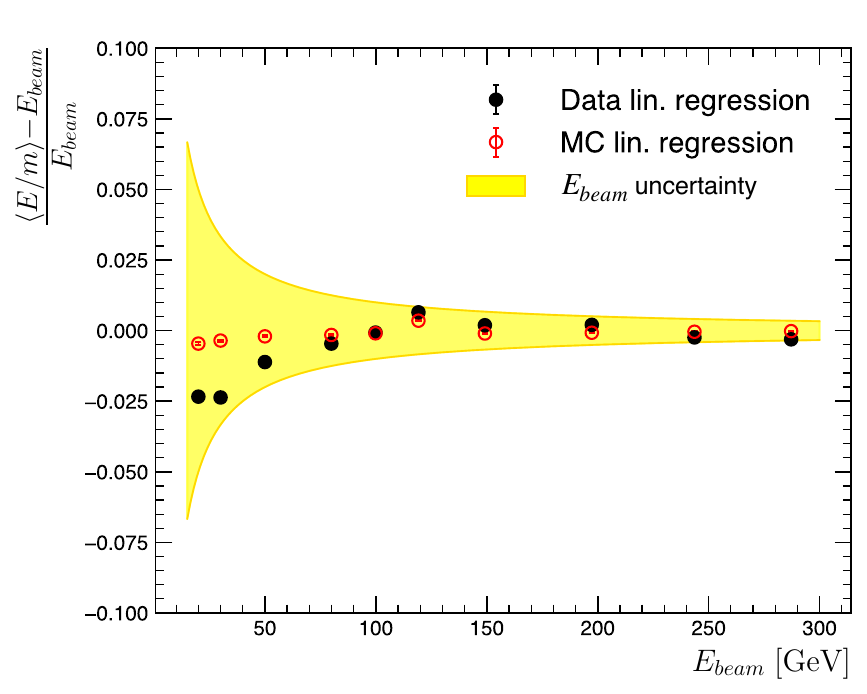}
  \caption{\label{fig:LinDataMCvis} Linearity with the energy. The measured energies $E$ are divided by the slope $m$ obtained from a linear fit to $<\!E\!>$ as a function of $E_{beam}$, with the slope and intercept allowed to float. The yellow band represents the relative error on the beam energy from the uncertainty of the dipole currents corresponding to an uncertainty of $\pm 1$ \GeV in momentum. 
  } 
\end{subfigure}
\end{figure*}

In Figure ~\ref{fig:ScaleFactor}, the mean measured energies in the data are compared to those in MC as a function of the beam energies.
The lower panel in Figure~\ref{fig:ScaleFactor}, shows that the data and simulation agrees at a 2\% level with no systematic dependence on the beam energy. As a scale factor of 1.035 has been applied to the data, this indicates that the measured energies are 3.5\% lower in data than in MC.
The difference can be attributed to either electronic-gain intercalibration, an inaccurate description of the detector layout in the simulation, or to the specific physics list used in \GEANTfour. Extensive checks ruled out other causes. 
In Figure~\ref{fig:LinDataMCvis}, the linearity with energy is shown, where the linearity was defined as the relative difference of the reconstructed positron energy with respect to the beam energy. 
The reconstructed positron energies, in GeV, were obtained from the measured energies in MIP units divided by a slope-factor, $m$, determined for both data and MC from a linear fit to the mean measured energy as a function of the beam energy, with the slope and intercept allowed to float.
The fit residuals are less than 1\% for data and 0.5\% for MC. 
Figure ~\ref{fig:LinDataMCvis} also displays the relative error on the beam energy from the uncertainty of the dipole currents (yellow band).
The linearity, without correcting for the losses upstream of the calorimeter, is better than 3\% for data and 1\% in the simulation.

\begin{figure*}[!t]
  \begin{center}
  \begin{minipage}[r]{0.45\textwidth}
      \vspace{0pt}
      \includegraphics[width=0.95\textwidth]{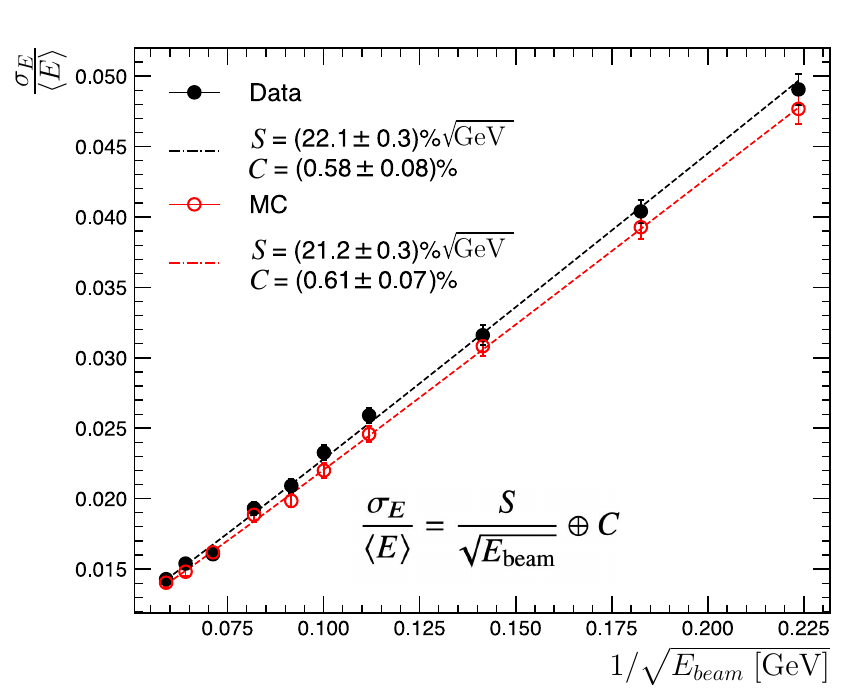}
      \vspace{-20pt}
      \end{minipage}
  \end{center}
  \caption{
    Relative energy resolution for measured energy in data and simulation. 
    }
  \label{fig:resolDataMCvis}
\end{figure*}

The values of the relative energy resolution,  
$\sigma_{E}/\left\langle E\right\rangle$, as a function of $1/{\sqrt{E_{\text{beam}}}}$, are shown in Fig.~\ref{fig:resolDataMCvis} for data and simulation, and are fit to the function:
\begin{equation}
  \frac{\sigma_{E}}{\left\langle E\right\rangle} = \frac{S}{\sqrt{E_{\text{beam}}}} \oplus C,  
        \label{eq:Fit_resol_funct}
\end{equation}
with $S$ the stochastic term and $C$ the constant term.
A noise term was not included in the fit because the residual contribution of the intrinsic noise is less than one MIP (see Section~\ref{Framework}) and is negligible after the hit selection chosen to reject noisy channels.
The results of the fits are shown in Fig.~\ref{fig:resolDataMCvis} and given in the two first columns of Table~\ref{tab06-2}.
Good agreement between data and MC for both the stochastic and constant terms is observed.
The constant term obtained in data is 0.6\% and is close to the value predicted by the simulation, indicating that energy response is uniform within the fiducial window defined by the DWC cut. 

So far, the previous results have been derived using only the sum of the measured energy in the Si pads. To study further optimization of the event reconstruction, two different methods to determine the energy of the positron were tested:
\begin{itemize}
\item
The Sampling Fraction (SF) method is based on the average sampling fraction over the complete detector~\cite{Paganis:2017def}, $SF(E_{\text{beam}})$, which is estimated with the simulation for each beam energy $E_{\text{beam}}$:
\begin{equation}
SF(E_{\text{beam}}) = \frac{ \sum_{i=1}^{28} E^{\text{Si}}_i}{
  \sum_{i=1}^{28}\left(E^{\text{Si}}_i+E^{\text{Abs}}_i \right)},
\label{eq:SFave}
\end{equation}
where $E^{\text{Si}}_i$ and $E^{\text{Abs}}_i$ are the energy deposited in the $i^{\text{th}}$ sensor and absorber.
The reconstructed energy deposited in the complete detector, $E$, is derived by applying the same energy dependent weight to all the layers:
\begin{equation}
 E = \frac{1}{SF(E_{\text{beam}})} \times \sum_{i=1}^{28} \left(E^{\text{Si}}_i[MIP] \times \Delta E_{i}^{\text{Si}}\right),
\label{eq:idealEtotSF}
\end{equation}
where, for the $i^{\text{th}}$ sensor,  $E^{\text{Si}}_i[MIP]$ is the measured energy in MIP units, and $\Delta E_{i}^{\text{Si}}$ is MIP value from simulation given in Table~\ref{tab:dEdxWeights}.
This method improves the linearity but does not alter the relative energy resolution. 
\item
The second method, the dEdx method, compensates for the energy losses in each absorber separately. In this, the energy deposited in the absorber layer $i$ is estimated as follows:
\begin{equation}
  n_{i}^{\text{Abs}} \times \Delta E_{i}^{\text{Abs}},  
  \label{eq:Abs_Energy_Theo}
\end{equation}
where $n_{i}^{\text{Abs}}$ is the energy deposited in the $i^{\text{th}}$ absorber expressed in terms of $\Delta E_{i}^{\text{Abs}}$, and $\Delta E_{i}^{\text{Abs}}$ is the mean minimum ionisation energy loss of a muon in the absorber given in Table~\ref{tab:dEdxWeights}, which was computed with the $dE/dx$ value of all the materials in the absorber \cite{PDG}.
Here, 
$n_{i}^{\text{Abs}}$ is estimated with the average of the 
measured energy in MIP units in the two sensors located before and after the absorber, 
and assuming only one charged incident particle in the first Si sensor:
\begin{linenomath*}
\begin{align}
  n_{1}^{\text{Abs}} &= \dfrac{1 + E_{1}^{\text{Si}}[\text{MIP}]}{2} \text{  and}\nonumber \\
  n_{i}^{\text{Abs}} &= \dfrac{E_{i-1}^{\text{Si}}[\text{MIP}] + E_{i}^{\text{Si}}[\text{MIP}]}{2} \text{   for } i=2,...,28.
  \label{eq:MIP_inAbs}
\end{align}
\end{linenomath*}
The combination of equations \ref{eq:Abs_Energy_Theo} and \ref{eq:MIP_inAbs}
results in an estimate of the total energy $E$ deposited in the passive and active layers of the prototype to be given by:
\begin{linenomath*}
\begin{align}
  E &=
  \frac{\Delta E_{1}^{\text{Abs}}}{2}+
  \sum_{i=1}^{27}
  \Big(
  \frac{\Delta E_{i}^{\text{Abs}} + \Delta E_{i+1}^{\text{Abs}}}{2} +\Delta E_{i}^{\text{Si}} \Big) \times E_{i}^{\text{Si}}[\text{MIP}]
  + \Big( \frac{\Delta E_{28}^{\text{Abs}}}{2} + \Delta E_{28}^{\text{Si}} \Big) \times E_{28}^{\text{Si}}[\text{MIP}] \nonumber \\
    &=   \frac{\Delta E_{1}^{\text{Abs}}}{2} + \sum_{i=1}^{28} \Big( W_i \mbox{  } E_{i}^{\text{Si}}[\text{MIP}] \Big).
  \label{eq:Abs_Energy_Rec_Detailed}
\end{align}
\end{linenomath*}
Since the layer dependent weights of the dEdx method, $W_i$, are independent of the beam energy, this methods hardly affects the linearity.
For the best pad intercalibration, the choice of the peak value was preferred over the mean and the most probable values of the muon energy loss distribution, with the difference for the dEdx method being only the overall energy scale.
\end{itemize}

For the SF method, there is less than 0.4\% difference between the sampling fraction value at 30\GeV and at 300\GeV \cite{CMS_NOTE}, 
so this method has almost the same linearity as found with just the measured energy. As expected, the SF method provides the same energy resolution as the resolution obtained from the measured energies.
The dEdx weights in equation \ref{eq:Abs_Energy_Rec_Detailed}, computed with Table~\ref{tab:dEdxWeights}, are identical except for the first and last layers, and for the layers 21 to 24.
Consequently, the dEdx method has almost the same energy linearity and resolution as that obtained with the measured energy.
The energy resolution stochastic and constant terms obtained with the two methods are summarized in Table~\ref{tab06-2}, showing a good agreement between data and MC.
\begin{table}[!t]
  \begin{center}
  \caption{
  The stochastic and constant terms $S$ and $C$ for data and simulation, defined in Eq.~\ref{eq:Fit_resol_funct} and obtained in the fit to the relative energy resolution for measured energies in Si, SF calibrated energies and dEdx calibrated energies.
    }
    \begin{tabular}{ |c||c|c||c|c||c|c| }
      \hline
      & \multicolumn{6}{|c|}{\textbf{Positron Energy Resolution}} \\
      \hline
      & Meas $S$                  & Meas $C$ & SF $S$                  & SF $C$& dEdx $S$                    & dEdx $C$  \\
      &\small$ [\sqrt{\text{GeV}}]$ \% & \small\%      & \small$ [\sqrt{\text{GeV}}]$ \% & \small\%      & $\small [\sqrt{\text{GeV}}]$ \% & \small\%      \\
      \hline
      & & & & & &\\
      Data & $22.1\pm 0.3$ & $0.58 \pm 0.08$ & $22.1 \pm 0.3$ & $0.58 \pm 0.08$ & $22.0\pm 0.3$ & $0.53 \pm 0.09$\\
      & & & & & &\\
      MC   & $21.2 \pm0.3$ & $0.61 \pm 0.07$ & $21.2 \pm 0.3$ & $0.61 \pm 0.07 $& $21.3 \pm0.3$ & $0.55 \pm 0.07$\\
      \hline
    \end{tabular}
    \label{tab06-2}
  \end{center}
\end{table}
\section{Position and Angular Resolutions}
\label{sec:positionres}
\newcommand{\pg}[0]{\vspace*{0.75em}\newline}
\newcommand{\percent}{$\%$}
The resolutions on the impact position and on the direction of the shower are relevant for a \mbox{particle-flow} optimized calorimeter. The results presented in this section were determined by using the track reconstructed with the DWCs as a reference for the trajectory of the incident particle.
In simulation, the DWC reconstructed track was obtained from a simple estimation of the DWC measurements  (see Section ~\ref{sec:datasimu}). 
\subsection{Position Reconstruction for Each Layer}
\label{subsec:pos_res_layer}
The reconstruction of the location of the centroid of the shower at each layer followed the procedure reported in \cite{HGCAL-2016TB} which is a logarithmic energy-weighted method:
\begin{equation}
  \label{eq:ImpactPosition}
  x_{\text{reco}} = \displaystyle \frac{\displaystyle \sum_{i \in M} \omega\left(E_{\text{pad }i}^{\text{Si}}\right) \cdot {x_{i}}}{\displaystyle
    \sum_{i \in M} \omega\left(E_{\text{pad }i}^{\text{Si}}\right)},~~\text{analogous for $y_{\text{reco}}$,}
\end{equation}
where $M$ included all selected hits within two rings of pads around the pad with the maximum deposited energy (a total of 19 pads).
The optimised energy weighting function $\omega\left(E_{\text{pad }i}^{\text{Si}}\right)$  was given by:
\begin{equation}
	\omega(E_{\text{pad }i}^{\text{Si}}) = \max\left[0,~a+\ln\left(\frac{E_{\text{pad }i}^{\text{Si}}}{\displaystyle \sum_{j \in M} E_{\text{pad }j}^{\text{Si}}}\right)\right]~,~~a = 3.5~~.
	\label{eq:logPosRecWeighting}
\end{equation}
The position residuals were defined as the difference in $x$ and $y$ between the reconstructed shower position and track extrapolation from the DWCs. 
For each layer and positron energy, the $x$- and $y$-position residuals were fit iteratively until convergence with a Gaussian function in the range of -2.0$\sigma$ to +2.0$\sigma$ around the mean
to extract the means and the standard deviations.

\begin{figure}[!ht]
  \centering
    \includegraphics[width=0.45\textwidth]{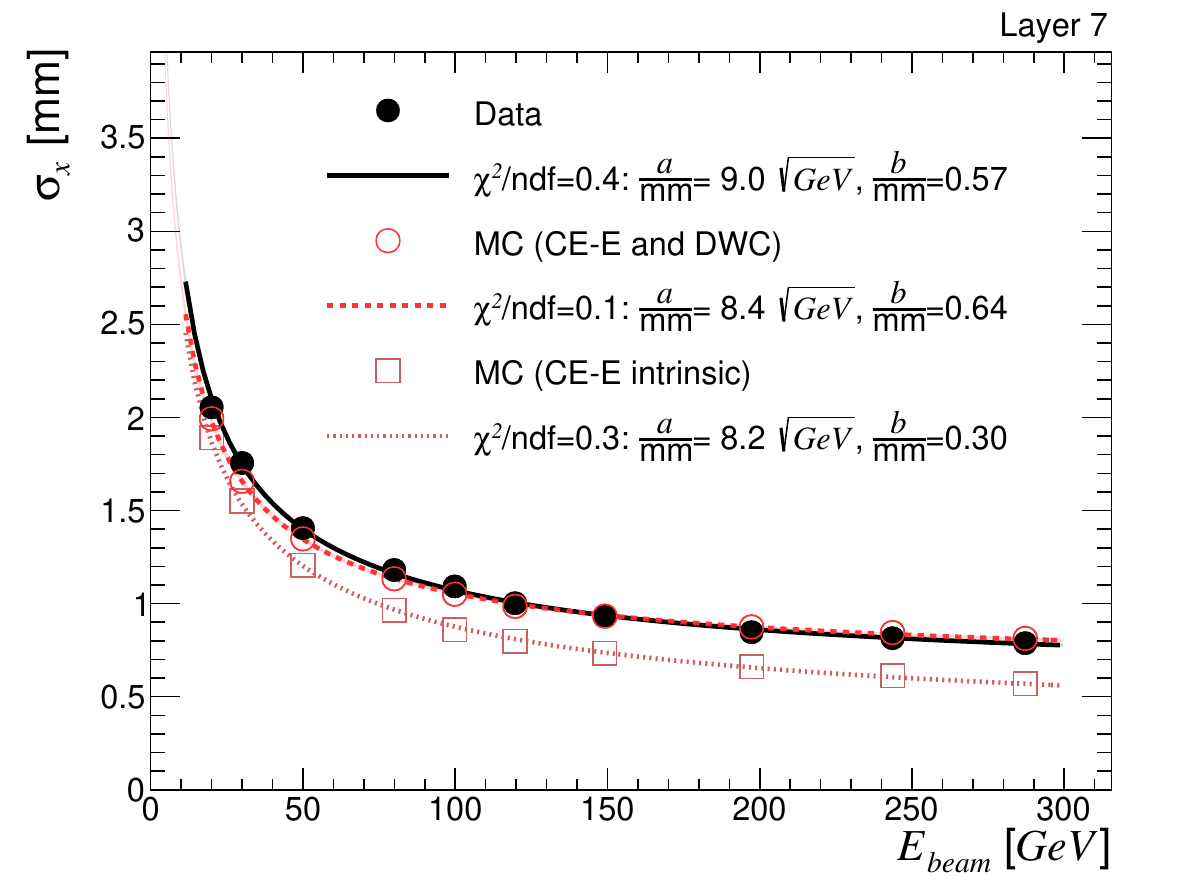}
  \caption{ 
    Combined CE-E prototype and DWC position resolution in $x$ in the single layer located at a depth of 6.7$~X_{0}$ as a function of the incident positron energy for data and simulations with (CE-E and DWC) and without (CE-E intrinsic) DWC resolution included in the DWC simulated measurements.
  }
  \label{fig:EMPosRes_2}
\end{figure}

The standard deviations served as a proxy for the position resolution, $\sigma_{x}$ and $\sigma_{y}$, which are the combination of the CE-E prototype and the DWC resolutions.
The position resolution in $x$ as a function of the beam energy in one layer (Layer~7) for data and simulation, is shown in Fig~\ref{fig:EMPosRes_2}. The intrinsic position resolution of the calorimeter, using the track position from the simulated DWC measurements without including the DWC resolution, is also shown.

The position resolutions were parameterized as a function of $E_{\text{beam}}$ using the function:
\begin{equation}
  \sigma_{x}~=~ \frac{a}{\sqrt{E_{\text{beam}}}} ~ \oplus ~b,~~\text{analogous for $y$},
  \label{eq:posResEnergy}
\end{equation}
where the stochastic term $a$ was motivated by the sampling fluctuations in the contributing Si pads. The results of the fits are displayed in Fig.~\ref{fig:EMPosRes_2} and given in Table~\ref{tab-xyposition}.
Below 100~\GeV, where multiple Coulomb scattering (MCS) is not negligible, a small difference between data and simulation including finite DWC resolution is observed and is probably due to the imprecision of the straight-line extrapolation used in the simulation of the DWC measurements (see Section~\ref{sec:datasimu}). 
For positron energies greater than 100~\GeV, the simulation including the finite DWC resolution reproduces well the data. Subtracting in quadrature the resolution obtained in the data from the intrinsic resolution obtained in the simulation is in good agreement with the DWC tracking resolution discussed in Section~\ref{sec:datasimu}.
In this energy range, the CE-E intrinsic position resolution obtained from the simulation, which is less than 0.8\mm, is a good estimate of the intrinsic resolution that can be obtained close to the electromagnetic shower maximum.
With a better tracking and a full beamline simulation, these results are consistent with those reported in~\cite{HGCAL-2016TB}.

\subsection{Reconstruction of the Shower Axis}
\label{subsec:pos_res_axis}
To estimate the shower axis, lateral shower impact positions in the data needed to be first corrected for any misalignment of the layers \cite{CMS_NOTE}.
The shower axis was determined from a straight line fit to the shower positions in consecutive layers beginning with the first layer that had more than 1\% of the total measured energy, and stopping when less than 5\% of the total energy remained.

The position residuals were defined as the distance in $x$ and $y$ between the reconstructed shower axis and the DWC track, evaluated at a depth equal to the shower longitudinal centre of gravity. 
The shower longitudinal centre of gravity, $COG_z$ in units of radiation length, was defined as:
\begin{equation}
COG_z[X_0] = 
\frac{ \sum_{i=1}^{28} E_i^{\text{Si}} \cdot z_{i}[X_0]}{ {\sum_{i=1}^{28}} E_{i}^{\text{Si}} },
\label{eq:05_cog}
\end{equation} 
where $E_i^{\text{Si}}$ is the energy deposited in the layer $i$ and $z_i[X_0]$ is the total radiation length up to the sensor $i$. 
The linear fits to the mean values of the $x-$ and $y-$position residuals as a function of $\Delta COG_{z}$ were used to estimate 
the relative angles, $\theta_x$ and $\theta_y$, between the CE-E prototype and the DWCs \cite{CMS_NOTE}.
$\Delta COG_{z}$ is the difference between the shower $COG_z$ for an event with respect to the  average value.
The resulting fit angles were found to be in the order of 10 mrad in both $x$- and $y$- directions with an estimated uncertainty of 0.7 mrad.
These relative angles were included in the simulation as the angles between the calorimeter and the beam.

\begin{figure*}[!ht]
\centering
\begin{subfigure}{.44\textwidth}
    \includegraphics[width=0.999\textwidth]{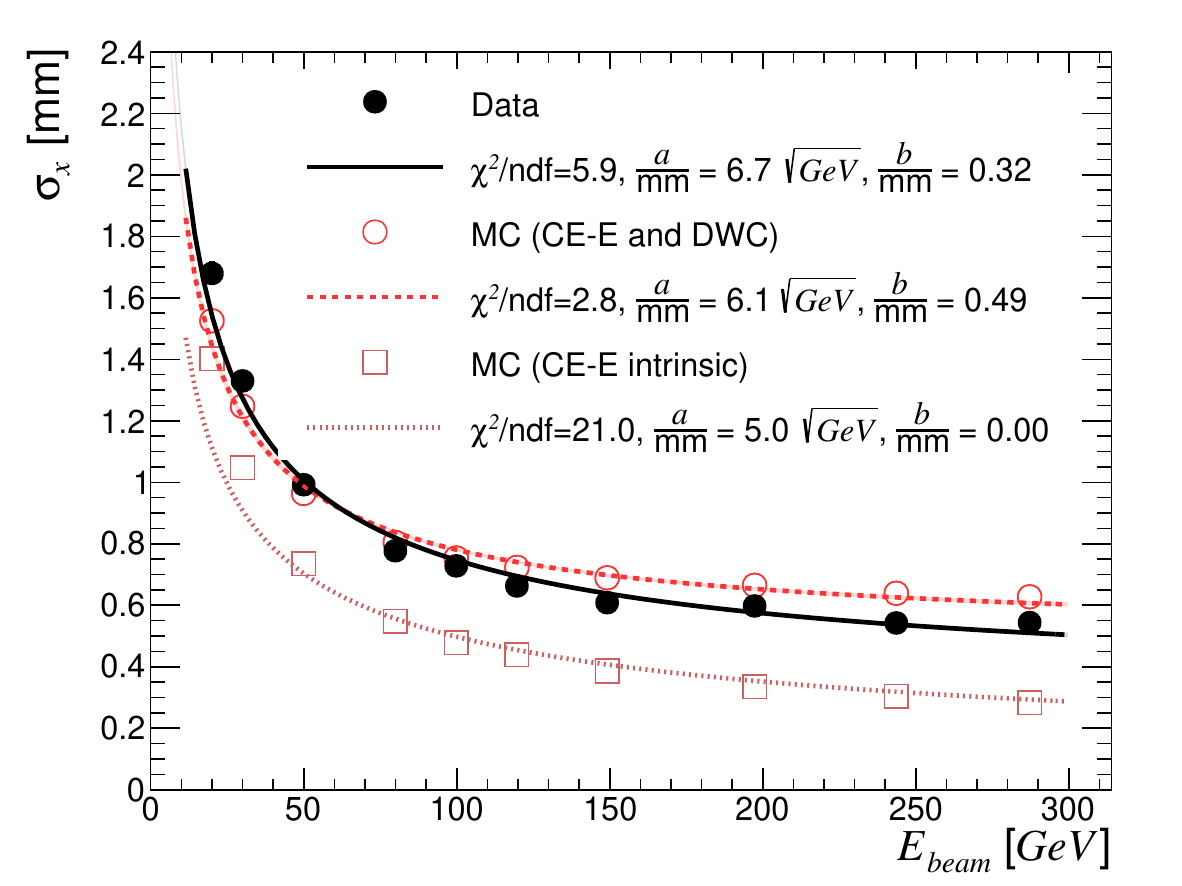}
    \caption{\label{fig:EMPosRes_4a} Combined shower axis and DWC position resolution at the $COG_z$ for data and simulations with (CE-E and DWC) and without (CE-E intrinsic) DWC resolution.} 
  \end{subfigure}
\qquad
\begin{subfigure}{.44\textwidth}
    \includegraphics[width=0.999\textwidth]{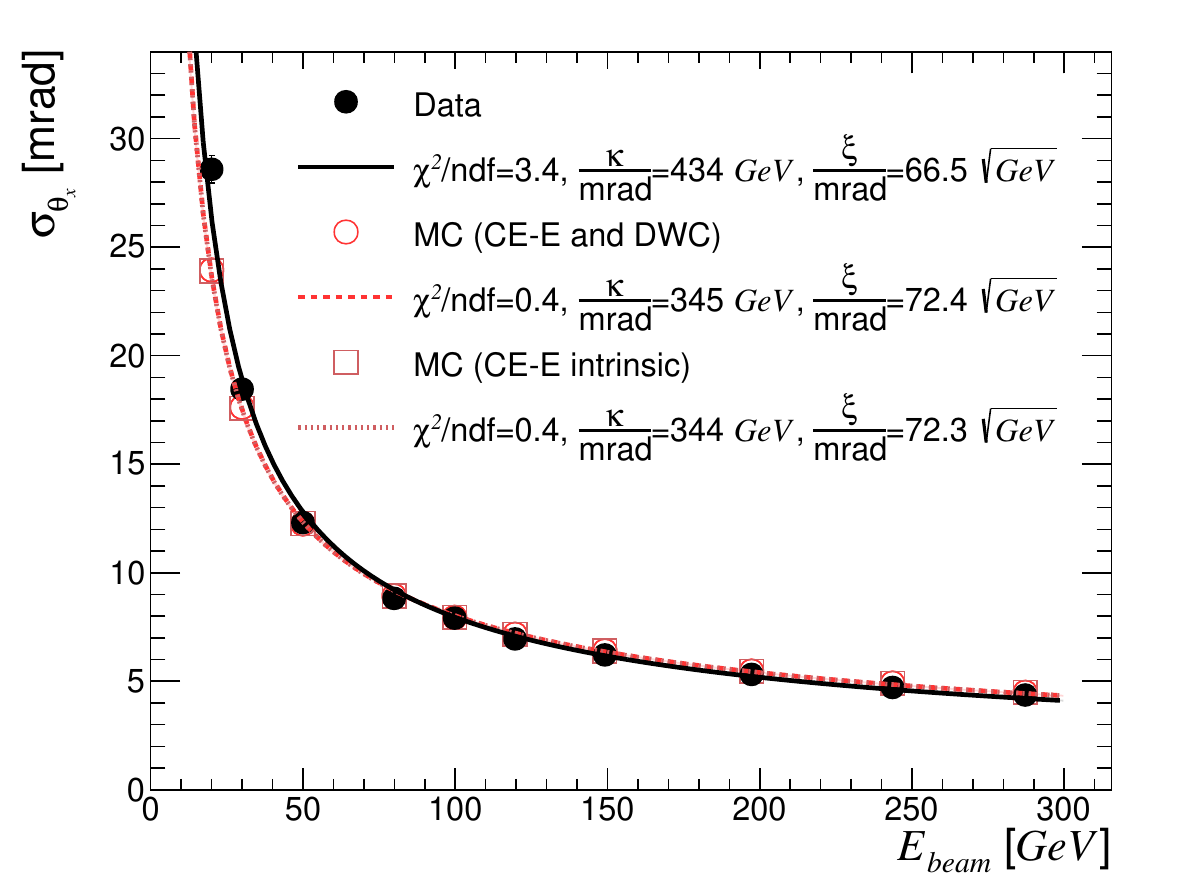}
  \caption{\label{fig:EMPosRes_4b} 
    Shower axis angular resolution for data and simulations with (CE-E and DWC) and without (CE-E instrinsic) DWC resolution. \newline
  }
\end{subfigure}
\end{figure*}

The combined shower axis and DWC position resolution in $x$ at the $COG_z$ is shown as a function of the beam energy in Fig.~\ref{fig:EMPosRes_4a}, and was fit to the function given in Eq.~\ref{eq:posResEnergy}. Results are given in Table~\ref{tab-xyposition}.
For the higher energies, there is a difference of the order of 0.1\mm between data and MC. This difference could be ascribed to an inaccurate description of the modules (structure of the PCB, pad geometry of the sensors) degrading the MC resolution, or to the rotational misalignment of modules in the prototype that could improve the data resolution.
At the highest energies, the intrinsic position resolution of the calorimeter was estimated to be less than 0.3\mm, following the same reasoning as in Section~\ref{subsec:pos_res_layer}.
In Fig.~\ref{fig:EMPosRes_4b}, the angular resolution as a function of the beam energy in the $z-x$ plane, $\sigma_{\theta_x}$, is shown. A similar performance was found in the $z-y$ plane.
The function in Eq.~\ref{eq:angularResEnergy}, with noise- and stochastic-like
terms $\kappa$ and $\xi$ as free parameters, models well the energy dependence, as seen in Fig.~\ref{fig:EMPosRes_4b}.
\begin{equation}
  \sigma_{\theta_x}~=~\frac{\kappa}{E_{\text{beam}}}~ \oplus ~\frac{\xi}{\sqrt{E_{\text{beam}}}},~~\text{analogous for $y$}.
  \label{eq:angularResEnergy}
\end{equation}
The values for $\kappa$ and $\xi$ are given in Table~\ref{tab-xyposition}. Data and simulation are in excellent agreement for energies larger than 30\GeV.
The effect of the position resolution of the DWC on the reconstructed track angle is negligible because of the 30~m lever arm between the first and last DWC.
In both data and simulation, the angular resolution at the highest energy point was found to be in the order of 4.5~mrad.
\begin{table}
  \begin{center}
  \caption{
    The stochastic and constant terms $a$ and $b$, defined in Eq.~\ref{eq:posResEnergy} and obtained in the fit to the combined CE-E prototype and DWC position resolution in $x$ in the layer located at a depth of 6.7 $X_0$. 
    The same terms $a$ and $b$ obtained in the fit to the combined shower axis and DWC position resolution in $x$ at the $COG_z$. 
    The noise and stochastic 
    terms $\kappa$ and $\xi$, defined in Eq.~\ref{eq:angularResEnergy} and obtained in the fit to the shower axis angular resolution in $x$. The simulation labelled MC$_\text{I}$ does not factor in the DWC resolution.
    }
    \begin{tabular}{ |c||c|c||c|c||c|c| }
      \hline
       & \multicolumn{4}{|c||}{\textbf{Position  Resolution}} & \multicolumn{2}{|c|}{\textbf{Angular  Resolution}}\\
      \hline 
      &One layer $a$                  & One layer $b$ & Axis $a$                  & Axis $b$& Axis $\kappa$                    & Axis $\xi$  \\
      & \small $[\mm \sqrt{\GeV}]$ & \small$[\mm]$  & \small$[\mm \sqrt{\GeV}]$ & \small$[\mm]$ & $[\text{mrad} \GeV]$ & \small$[\text{mrad} \sqrt{\GeV}]$\\
      \hline
      & & & & & &\\
      Data           & $9.0\pm 0.2$ & $0.57 \pm 0.02$ & $6.7 \pm 0.1$ & $0.32 \pm 0.02$ & $434 \pm 12$ & $66.5 \pm 1.0$\\
      & & & & & &\\
      MC  & $8.4 \pm0.2$ & $0.64 \pm 0.02$ & $6.1 \pm 0.1$ & $0.49 \pm 0.02 $& $345 \pm 13$ & $72.4 \pm 0.9$\\
      MC$_\text{I}$  & $8.2 \pm0.2$ & $0.30 \pm 0.02$ & $5.0 \pm0.1$ & $0.00 \pm 0.02 $& $344 \pm 13$ & $72.3 \pm 0.9$\\
      \hline
    \end{tabular}
    \label{tab-xyposition}
  \end{center}
\end{table}
\section{Shower Shape Measurements}
\label{sec:showershape}
%
%
The longitudinal and lateral development of the electromagnetic showers obtained with the fine sampling of the CE-E prototype are presented here and compared to the simulation. The validity of an empirical parameterization of the longitudinal profile and the crosstalk effects on the lateral shower shapes are also discussed.

\subsection{Longitudinal Shower Shapes}
\begin{figure*}[ht!]
  \begin{center}
    \includegraphics[width=0.35\textwidth]{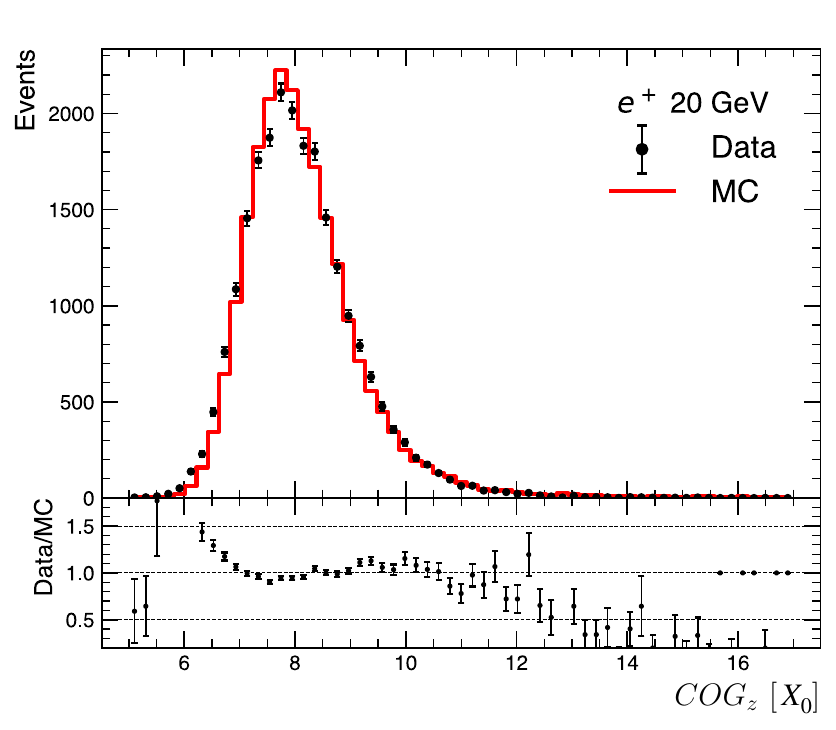}
    \includegraphics[width=0.35\textwidth]{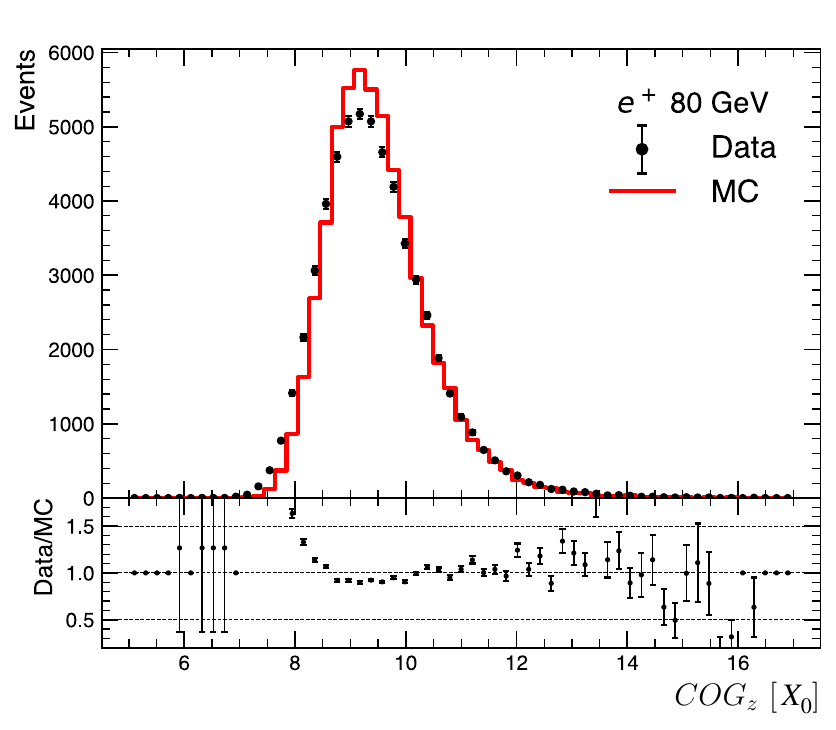}
    \includegraphics[width=0.35\textwidth]{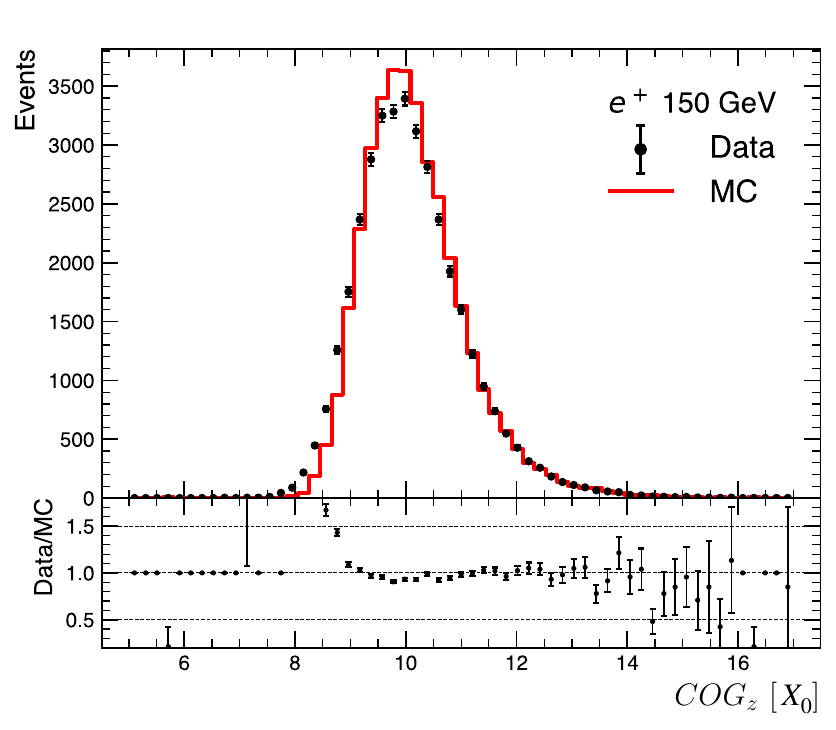}
    \includegraphics[width=0.35\textwidth]{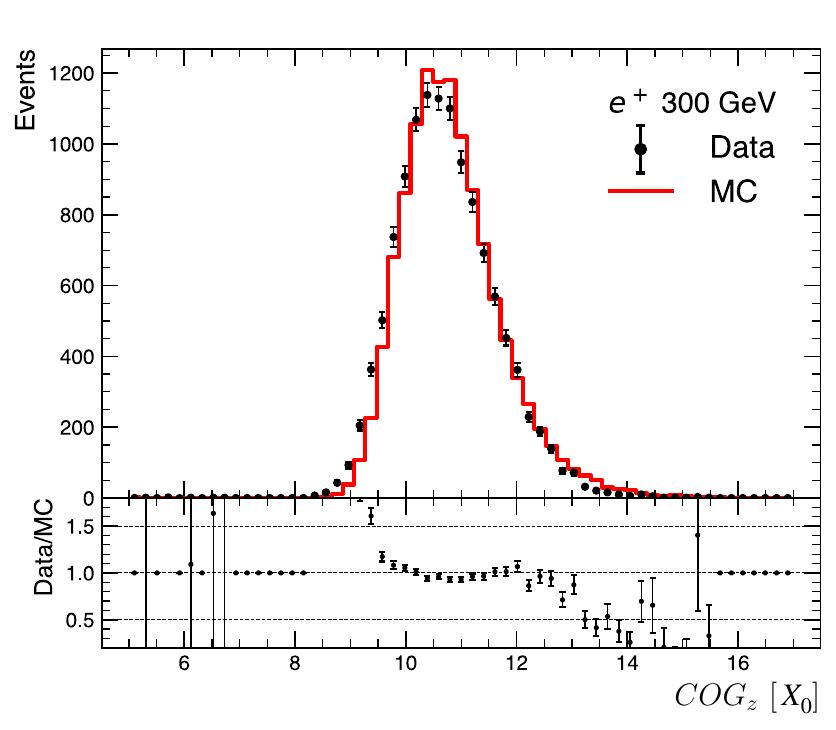}
    \caption{Distributions of shower longitudinal centre of gravity  for nominal positron beam energies ranging from 20 to 300~\GeV. The simulation is normalized to the number of events in data.
    }
    \label{fig:long-01}
  \end{center}
\end{figure*}

Comparisons between data and simulation of the shower longitudinal centre of gravity distributions are shown in Fig.~\ref{fig:long-01}, for nominal positron energies in the range 20 to 300~GeV.
A reasonable agreement between data and simulation is observed for the full energy range. 
The slight shift toward greater $X_0$ in the simulation can be attributed to the incomplete modeling of the H2 beamline.

\begin{figure}[!ht]
  \begin{center}
    \includegraphics[width=0.45\textwidth]{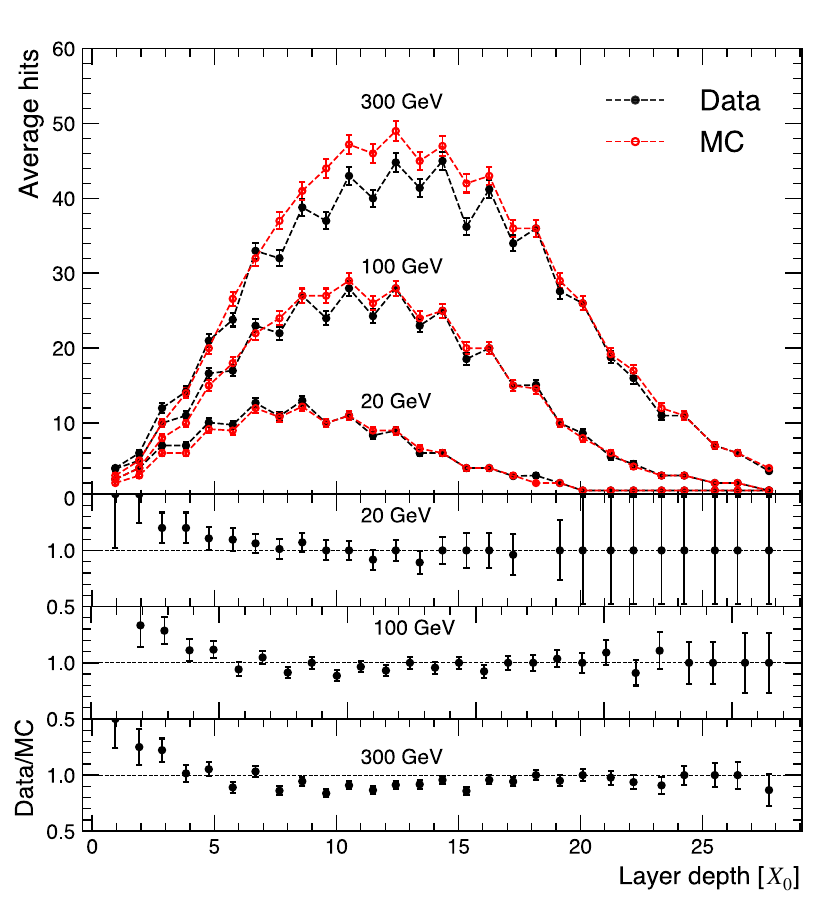}
    \includegraphics[width=0.46\textwidth]{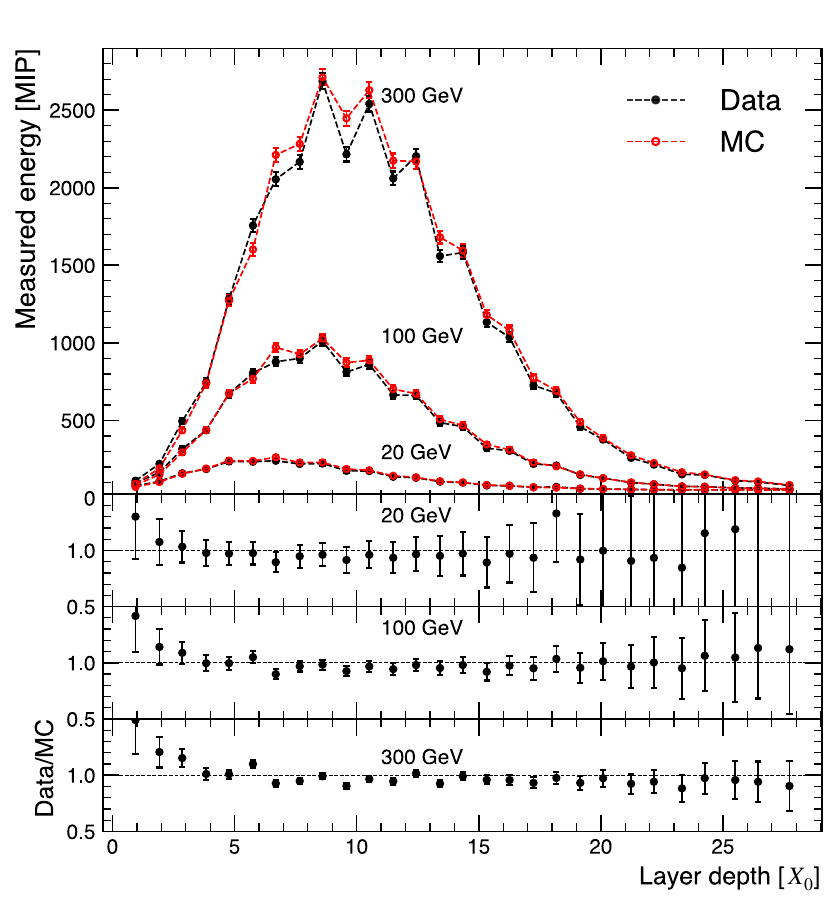}
    \caption{Longitudinal shower profiles, for different nominal positron beam energies, given by the average number of selected hits (left) and the median measured energy without applying any scale factor to the data (right). The ratio of the data to the MC point values are displayed in the lower panels (error bars are evaluated by propagating the errors on the values).
    }
    \label{fig:Long_profile}
  \end{center}
\end{figure}

In Fig.~\ref{fig:Long_profile}, the longitudinal profiles, given by the average number of selected hits and the median measured energy as a function of the layer depth, are shown for nominal positron beam energies of 20, 100 and 300~GeV.
An overall agreement with the simulation over the entire detector for the two types of profiles, is observed. 
Apart from the first layers, there is less energy measured in data than in simulation resulting in about 3.5\% less energy measured in the complete detector in data compared to MC, as outlined in Section~\ref{sec:energyrl}.
The average number of hits also displays fewer hits for data than MC in the central layers.
Despite the fact that for the absorbers, the mean minimum ionisation energy lost in even layers is close to double the energy lost in odd layers, the measured energy of the shower is higher in odd layers than in even layers. This effect, observed in both data and MC, was studied in simulated showers and attributed to two causes in relation to the cassette layout~\cite{CMS_NOTE}.
The larger energy deposited in odd Si sensors is due to a larger number of soft electrons, dominated by delta rays, produced in the material (especially the PCB) located directly in front of the odd-layer Si sensors, and additionally to a larger backward-moving soft electromagnetic component due to the CuW plate located right after the odd-layer Si sensors, compared to the even-layer sensors which were preceded by the CuW plate and followed by the PCB. 
The minor difference in the odd/even response between data and simulation is not understood.

\begin{figure*}[!ht]
\centering
\includegraphics[width=0.45\textwidth]{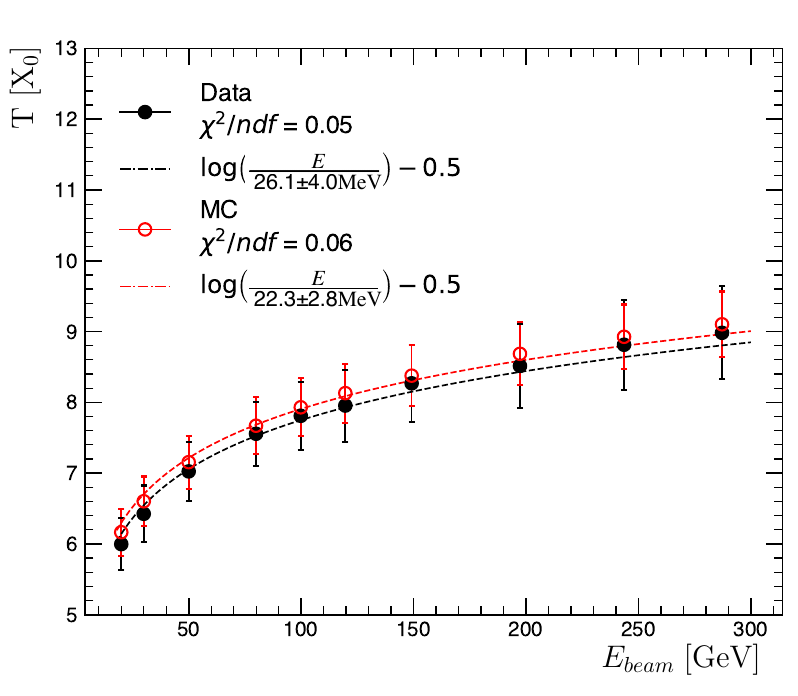}%
\caption{Shower maximum as a function of the beam energy. The shower maximum is determined from the longitudinal profile fit (error bars are evaluated by propagating the parameter fit uncertainties).
}
\label{fig:CriticalEnergy}
\end{figure*}

The average longitudinal electromagnetic shower profile for homogeneous media can be described empirically using the pa\-ra\-me\-te\-ri\-zation \cite{Longo:1975wb}:
\begin{equation}\label{Longo}
\left\langle \frac{dE(z)}{dz} \right\rangle = E_0 \frac{(\beta z)^{\alpha - 1} \beta \exp (- \beta z)}{\Gamma (\alpha )},
\end{equation}
where $z$ is the depth in radiation lengths, $E_0$ is the mean of the total energy deposited in the calorimeter and $\alpha $ and $\beta$ are the shape and scaling parameters, respectively. The analytical expression in Eq.~\ref{Longo}  gives a good first-order approximation for sampling calorimeters. 
The measured energy longitudinal profiles were fitted  with Eq.~\ref{Longo}, to examine the expected logarithmic energy dependence of the position of the longitudinal shower maximum $T$ given by $\frac{\alpha - 1}{\beta}$.
Figure~\ref{fig:CriticalEnergy} gives the shower maximum as a function of the beam energy. It was fitted using the parameterization \cite{PDG}:
\begin{equation}
T =  \log\left( \eta \right) - 0.5,
\label{Tmax_log}
\end{equation}
where $\eta  = E/E_c$ and $E_c$ is the critical energy of the calorimeter. The critical energy represents the electron or positron energy for which the ionisation and excitation losses are equal to those from radiative processes (bremsstrahlung and pair creation).
In all cases, fit residuals at the level of 4 to 5$\%$ were obtained with no dependence on beam energy, demonstrating the validity of the $\log (\eta )$-dependence of $T$. 
The $\chi^2/ndf$ values are especially small because point-to-point correlations, due to the odd/even difference of response observed in the longitudinal profiles, were not taken into account. 
The critical energy obtained is 26.1 $\pm$ 4.0 \MeV for data and 22.3 $\pm$ 2.8 \MeV in the simulation. Detailed studies are found in~\cite{CMS_NOTE} where the validity of the  parameterization is tested using the average $COG_z$, either extracted from the longitudinal fit or from the event distribution.

\subsection{Transverse Shower Shapes}
\begin{figure*}[!htb]
  \begin{center}
    \includegraphics[width=0.45\textwidth]{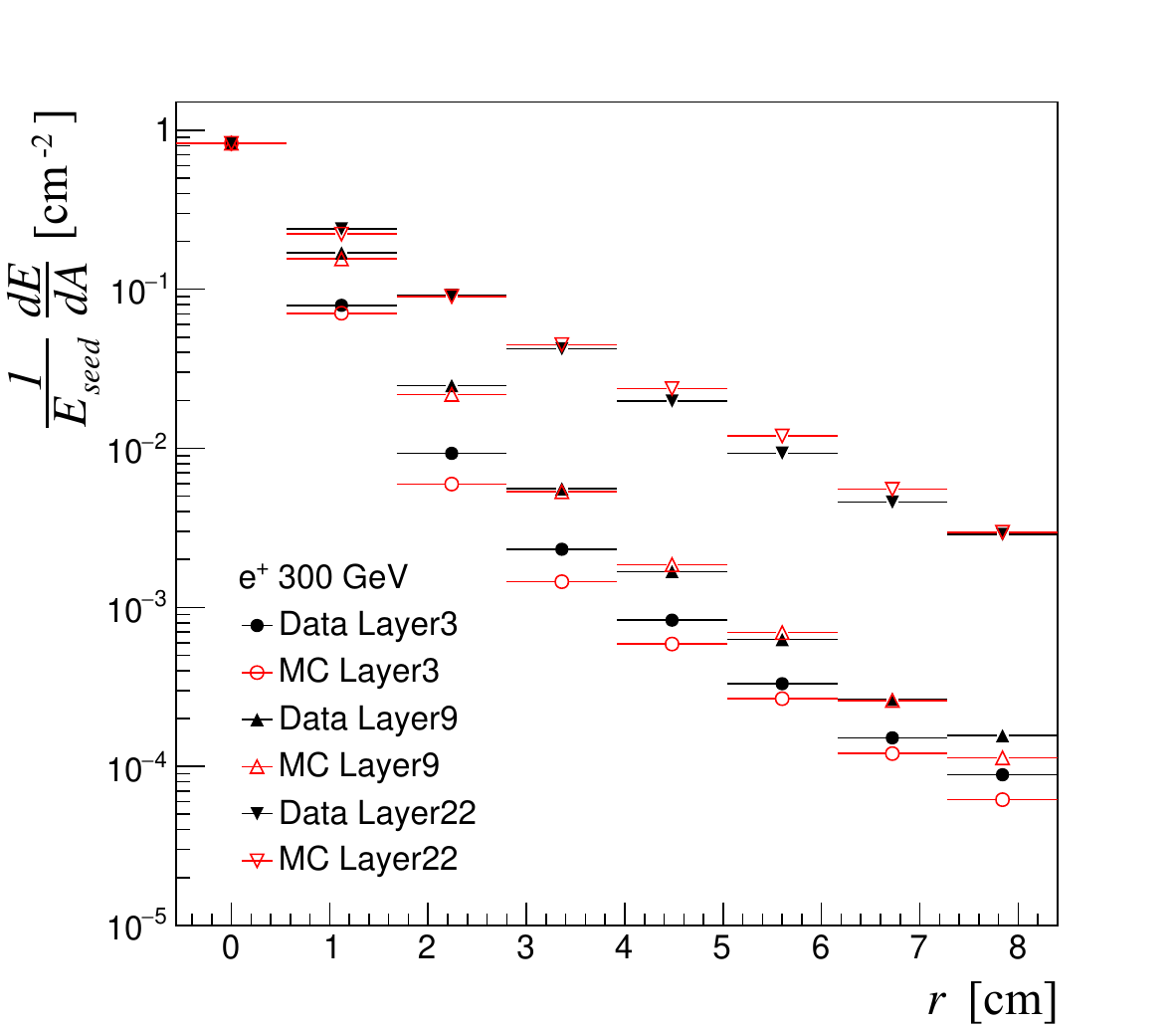}
    \caption{Transverse shower profile comparison between data and simulation for nominal positron energy of 300~\GeV in layers 3, 9 and 22. The shapes shown are the energy deposited in a ring of pads per units of active area as a function of its radial distance $r$ from the center of the seed pad, normalized to the energy of the seed pad in the layer under consideration.}
    \label{fig:transv-01}
  \end{center}
\end{figure*}

To study the lateral spread of electromagnetic showers in the CE-E, the "seed" pad of the energy spatial distribution in each layer was defined as the one with the maximum energy.
The transverse shower profile of a layer was determined as 
the energy deposited in a ring of pads per unit of active area as a function of the radial distance $r$ from center of the seed pad, normalized to the energy of the seed pad.
This representation allows, in principle, the parameterization of the lateral energy deposition as a function of shower depth and beam energy.
In Fig.~\ref{fig:transv-01}, the comparison of transverse shower profiles between data and simulation is shown for  300~GeV positrons, at three different depths. As expected, the energy density is steeper in the first layers.
A reasonable agreement between data and simulation is observed, particularly for the most energetic pads (those nearest to the seed pad).

\begin{figure*}[!htb]
  \begin{center}
    \includegraphics[width=0.27\textwidth]{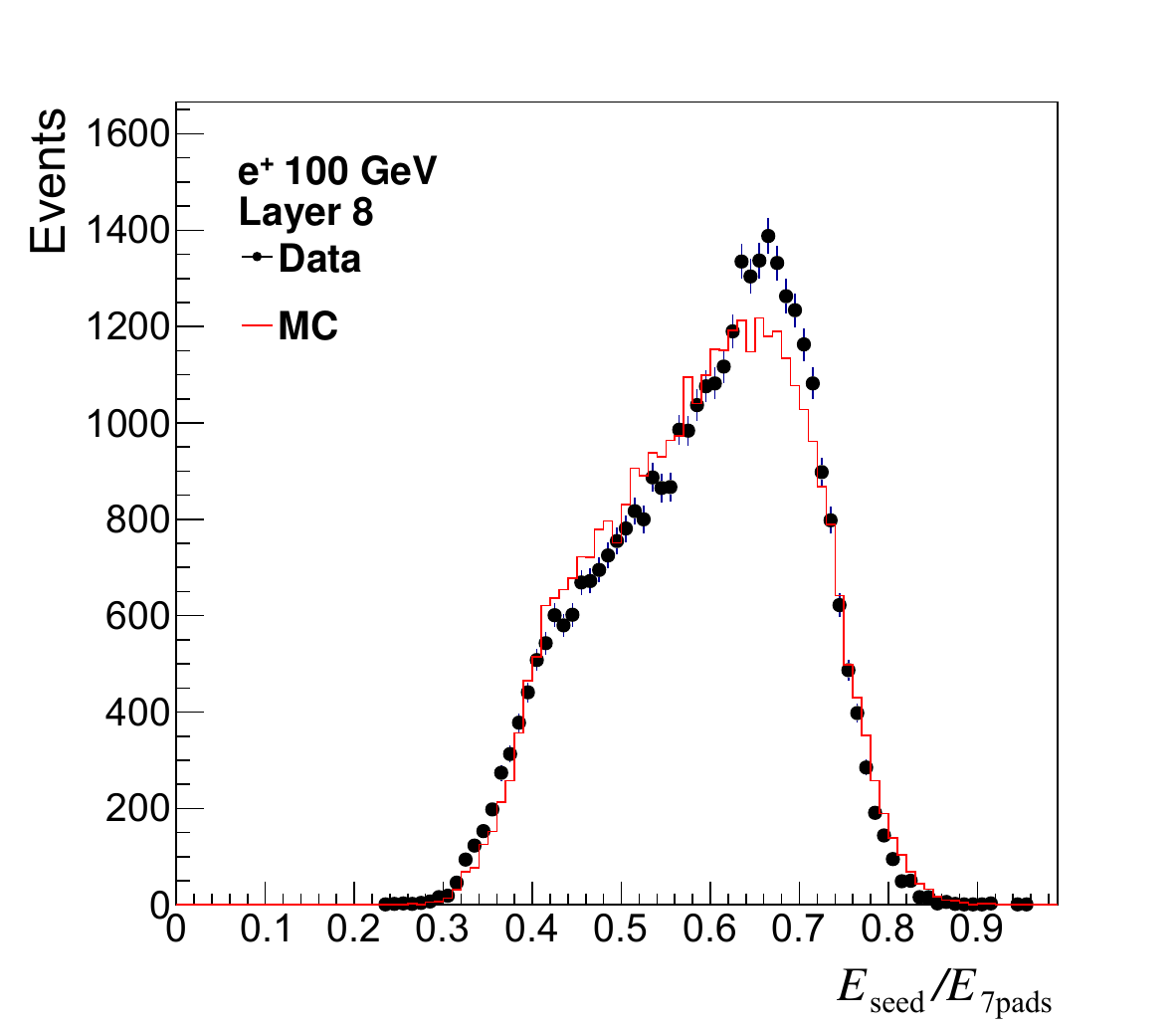}
    \includegraphics[width=0.27\textwidth]{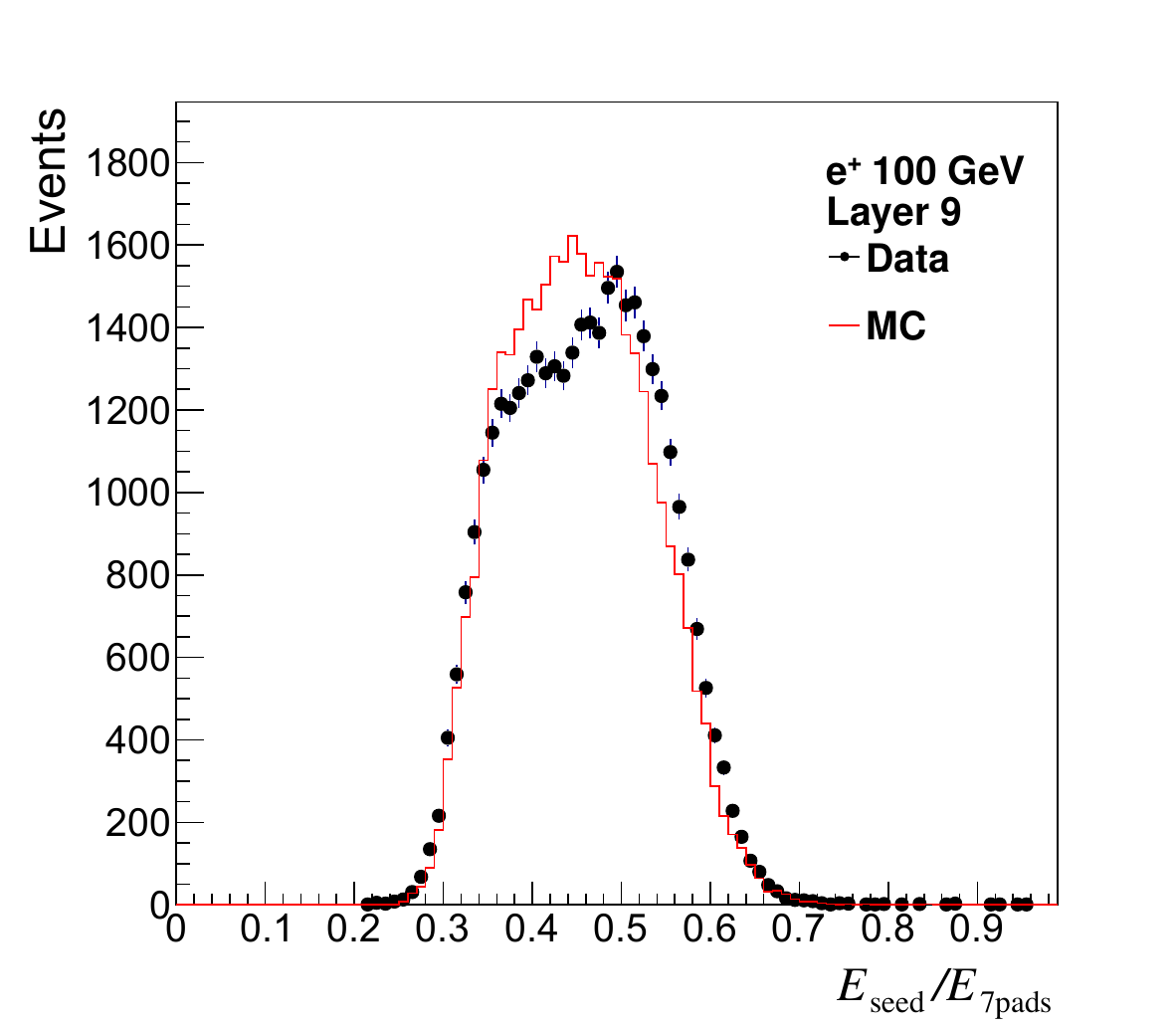}
    \includegraphics[width=0.27\textwidth]{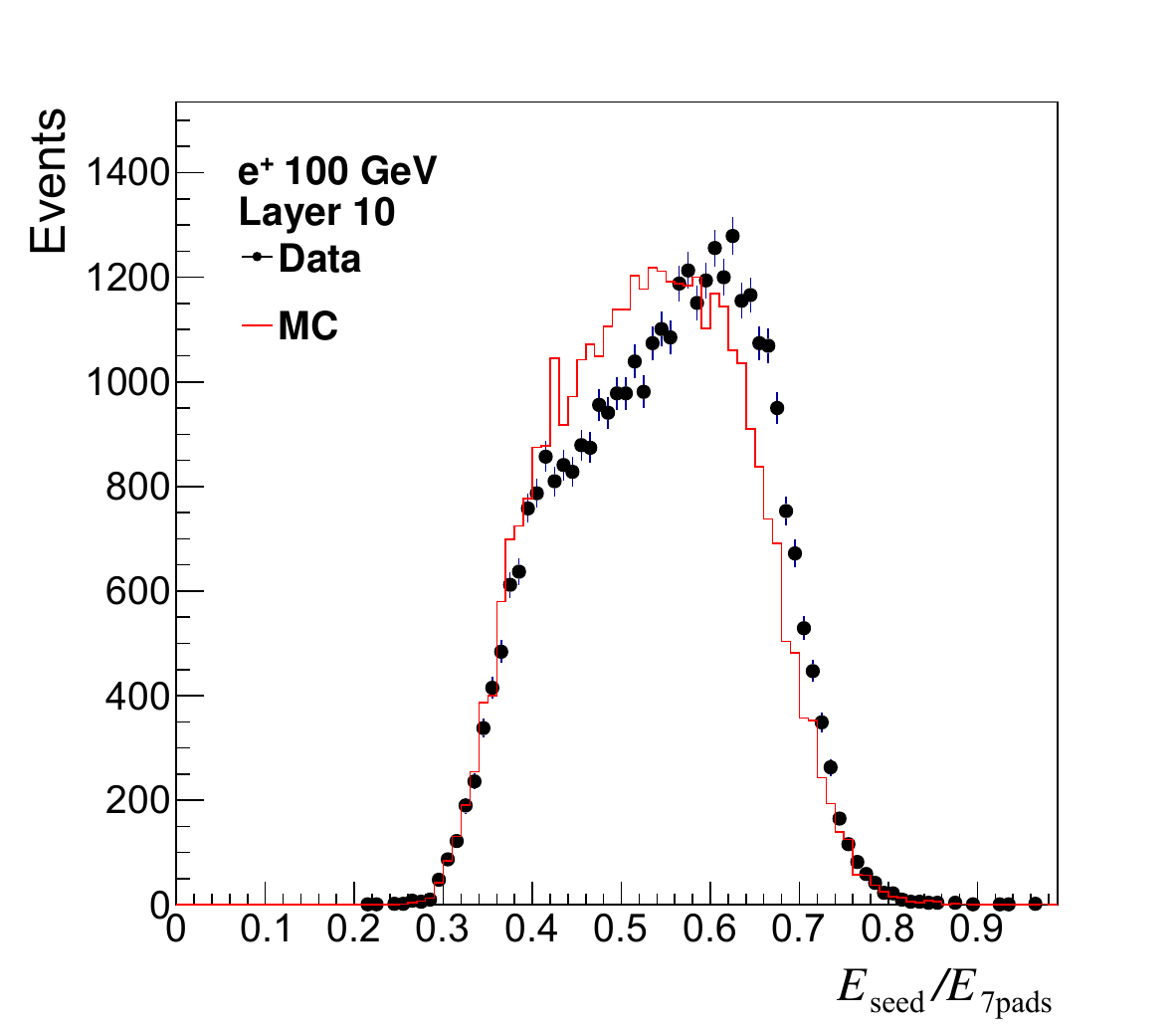}
    \includegraphics[width=0.27\textwidth]{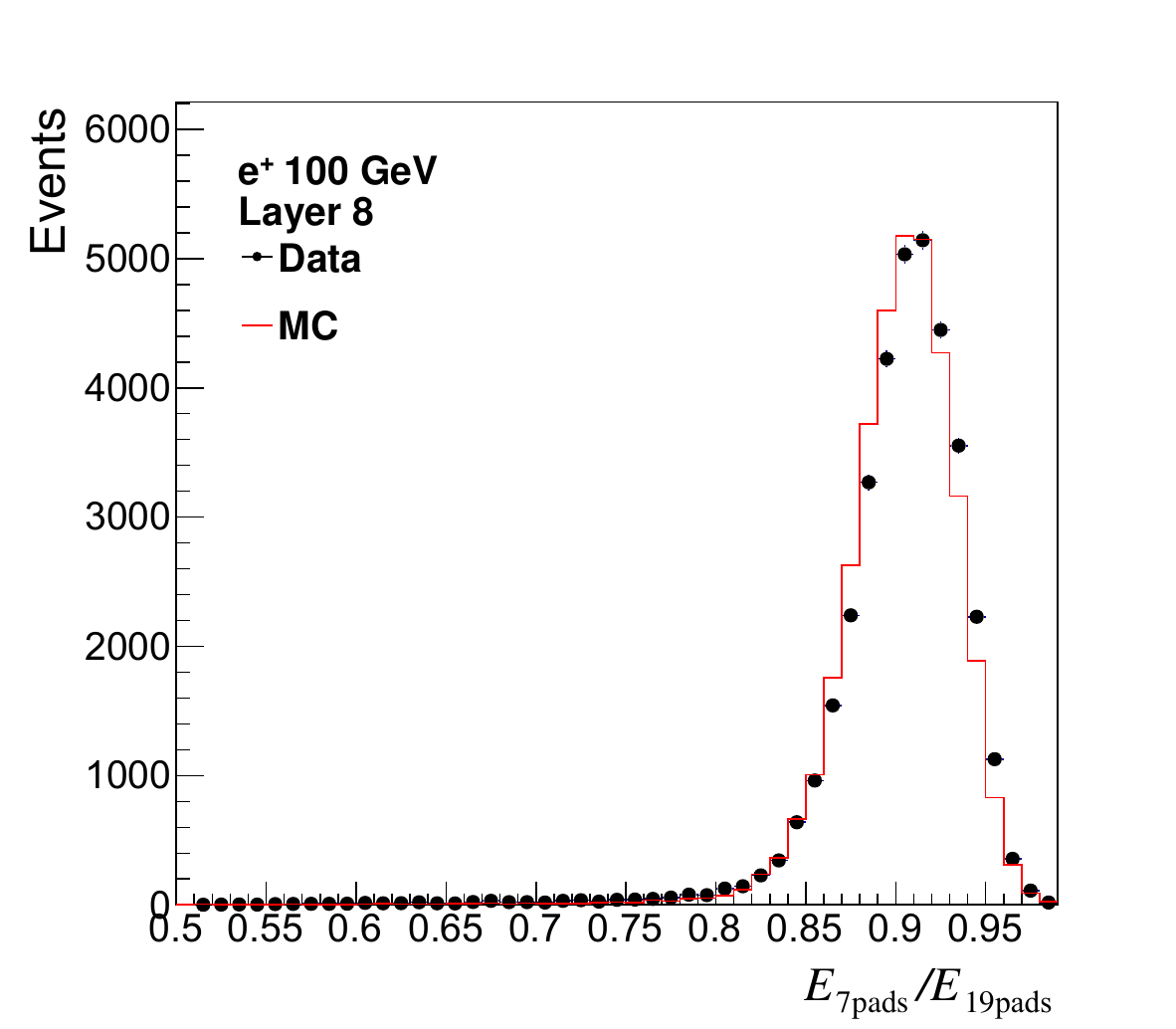}
    \includegraphics[width=0.27\textwidth]{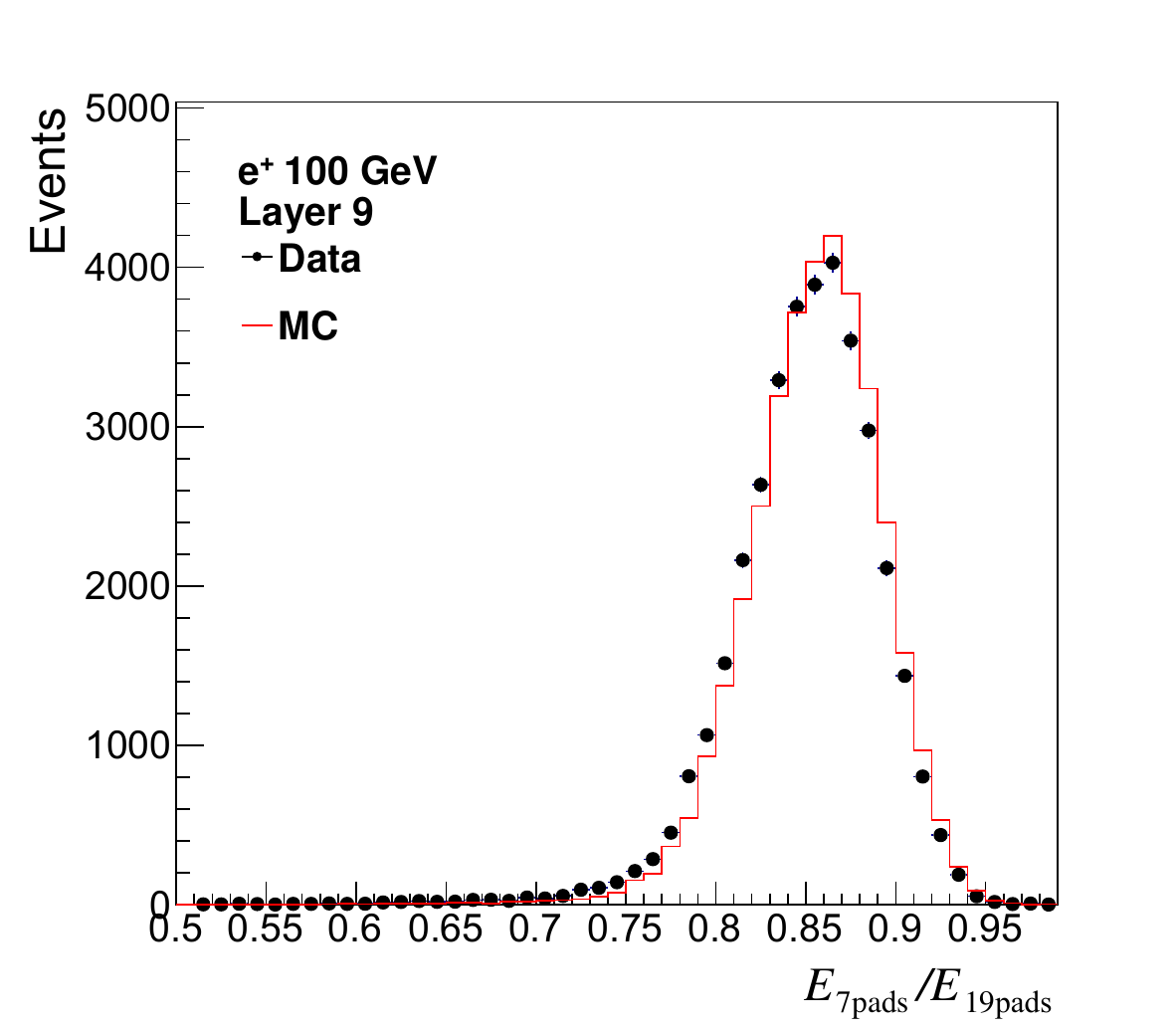}
    \includegraphics[width=0.27\textwidth]{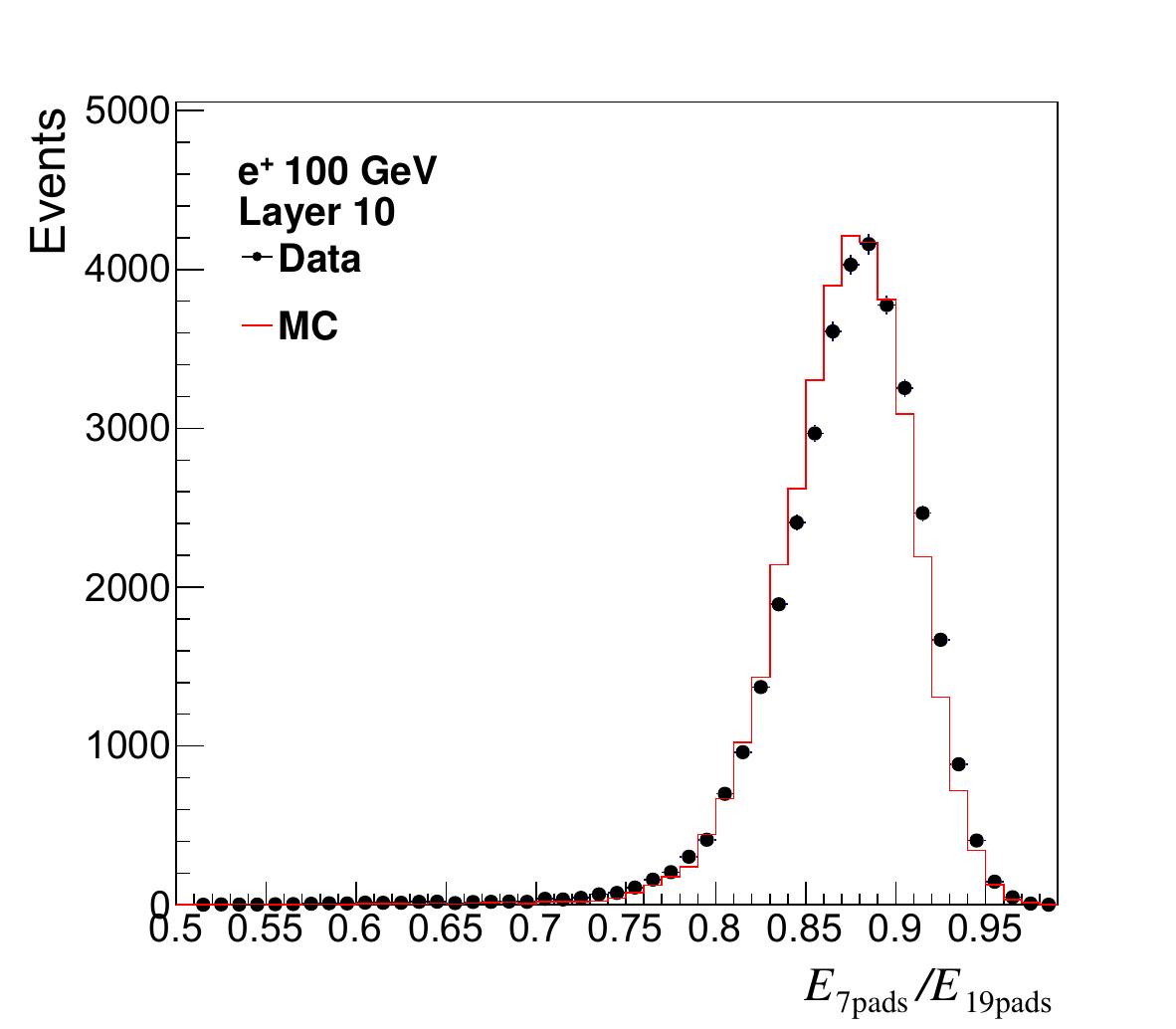}
    \caption{Distributions of $E_{\text{seed}}/E_{\text{7pads}}$ and $E_{\text{7pads}}/E_{\text{19pads}}$ for layers 8, 9 and 10 and for nominal positron energy of 100~\GeV.
    The simulation is normalized to the number of events in data.}
    \label{fig:e1oe7_e1oe19_setupI}
  \end{center}
\end{figure*}

Another method to study the transverse shower development is to find, for each layer, the ratio between the energy deposited in the seed pad and the 7 pads around it, $E_{\text{seed}}/E_{\text{7pads}}$, 
or the ratio between the energy deposited in 7 pads and the surrounding 19 pads, $E_{\text{7pads}}/E_{\text{19pads}}$.  
Figure~\ref{fig:e1oe7_e1oe19_setupI} shows these distributions in layers 8, 9 and 10 for a nominal positron energy of 100~GeV.
The $E_{\text{seed}}/E_{\text{7pads}}$ distributions  appear to be slightly shifted toward higher values in data compared to simulation, whereas the $E_{\text{7pads}}/E_{\text{19pads}}$ distributions display a small odd/even layer dependent shift.
Good agreement between data and MC in the constant term of the energy resolution indicates that pad-to-pad response variations (after pad intercalibration) present in data, are negligible and are not responsible for the shift between the data and the simulation in the $E_{\text{seed}}/E_{\text{7pads}}$ distributions. However, in some layers, isolated pads 
, which were not well calibrated,
might contribute to the shape of $E_{\text{seed}}/E_{\text{7pads}}$ in data without affecting the energy resolution. This is the case of layer 9, where the data shows a gap in the distribution. The beamline simulation provided a beam position spread in good agreement with the data. However, 
the distribution of the spatial-positron track impact on the calorimeter is slightly different between data and simulation because the mean beam impact position is not perfectly reproduced by the simulation. This is understood to be the main cause of the different $E_{\text{seed}}/E_{\text{7pads}}$ shapes in data and MC. Differences between data and simulation in the hadron contamination, or the bremsstrahlung missing energy due to upstream material, 
could also induce similar discrepancies in $E_{\text{seed}}/E_{\text{7pads}}$.

\begin{figure*}[!hbt]
  \begin{center}
    \includegraphics[width=0.3\textwidth]{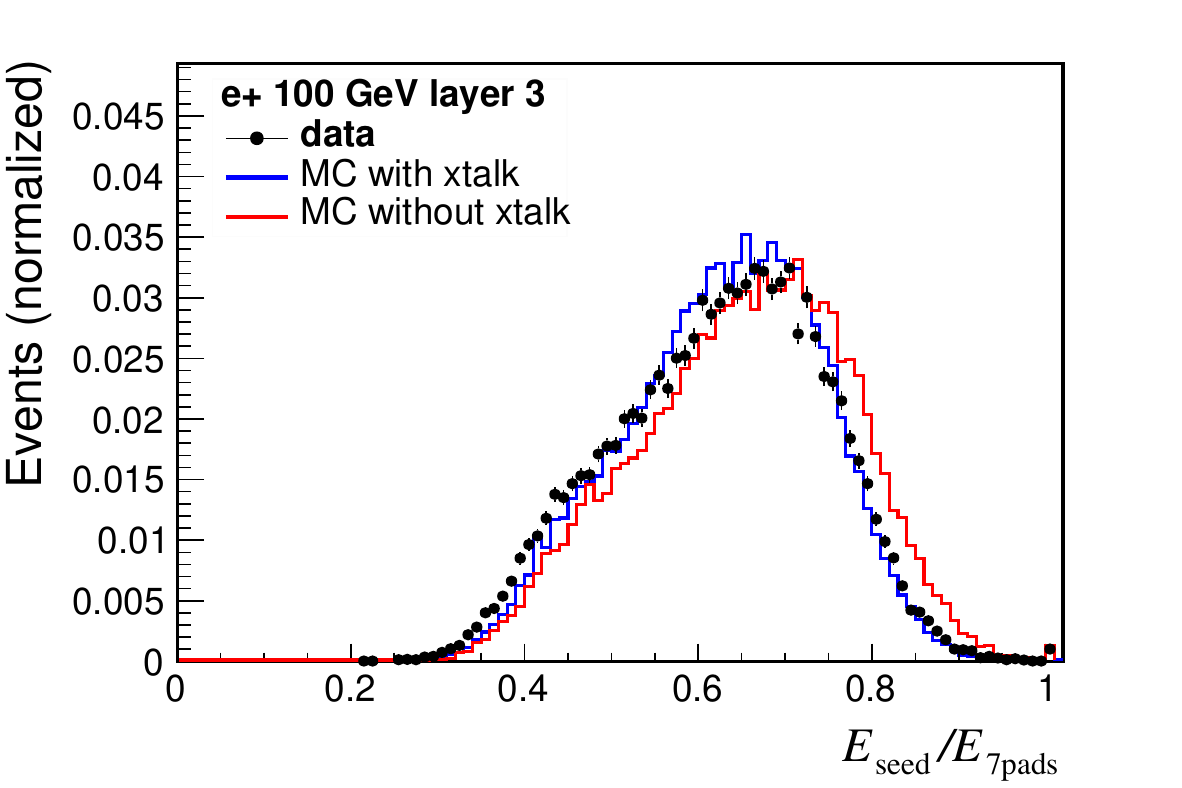}
    \includegraphics[width=0.3\textwidth]{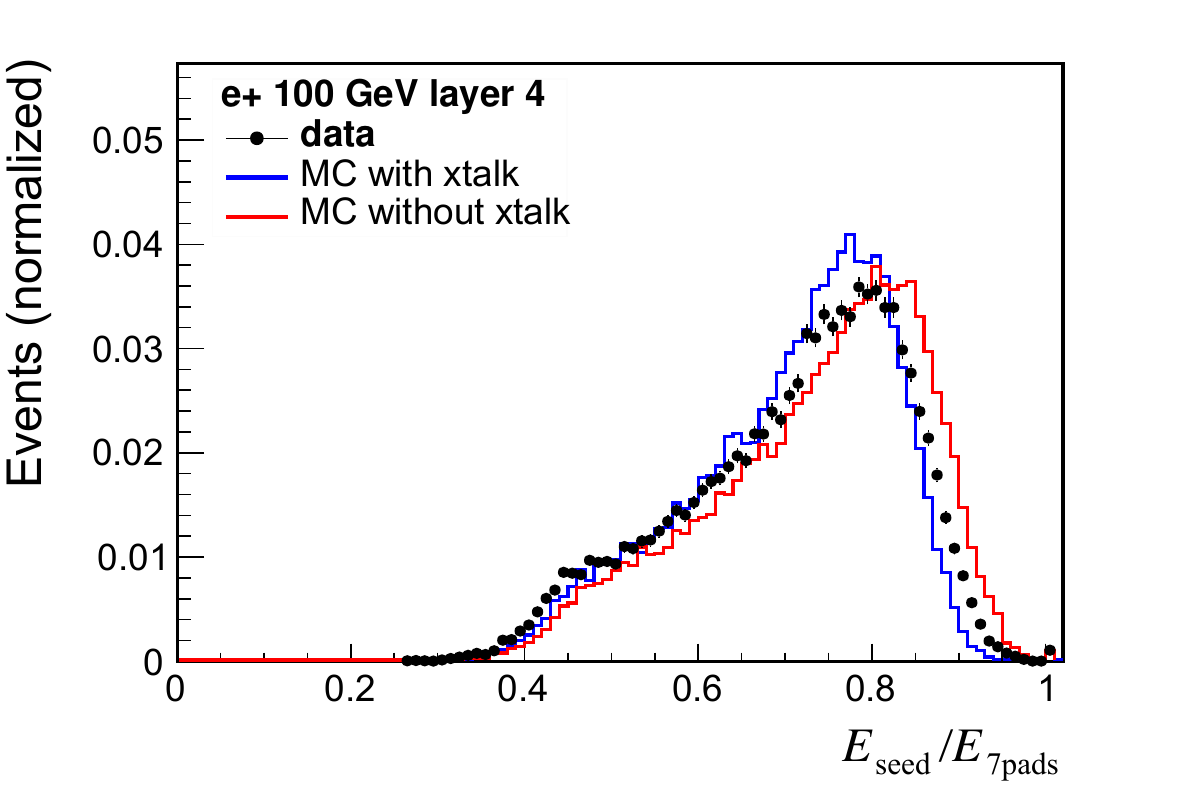}
    \includegraphics[width=0.3\textwidth]{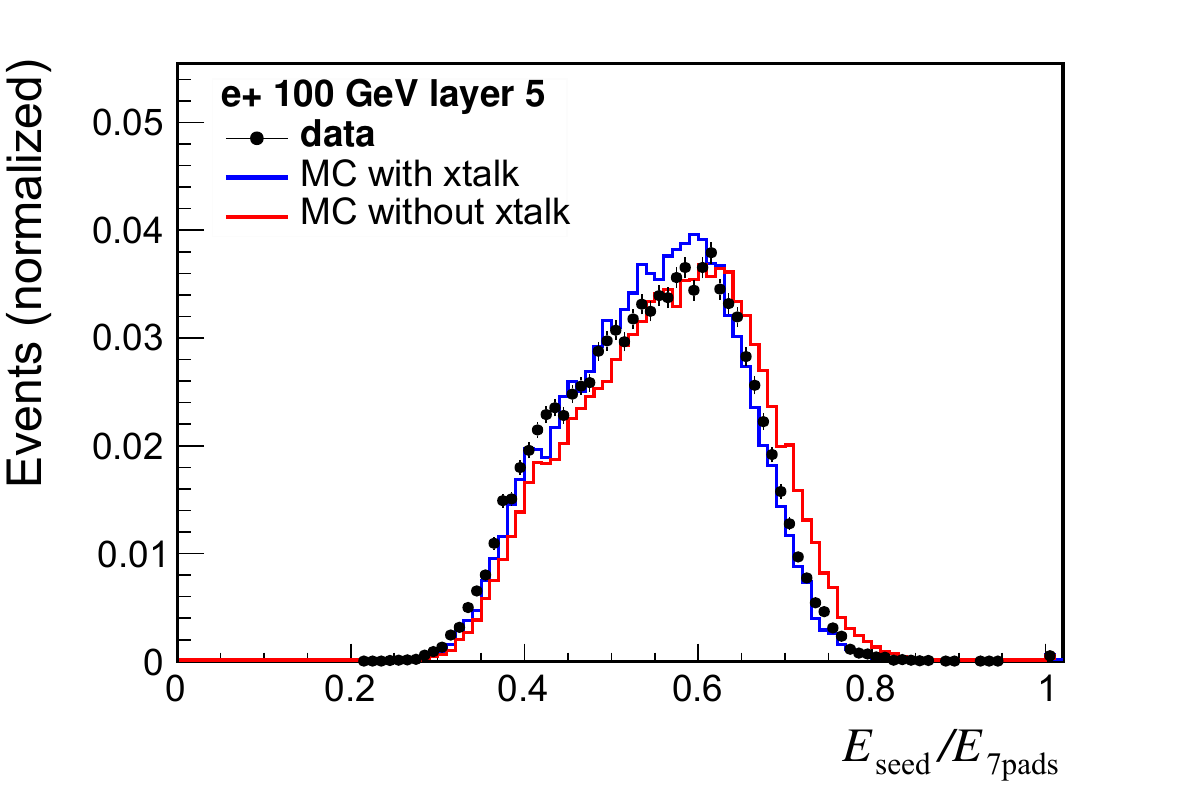}
    \includegraphics[width=0.3\textwidth]{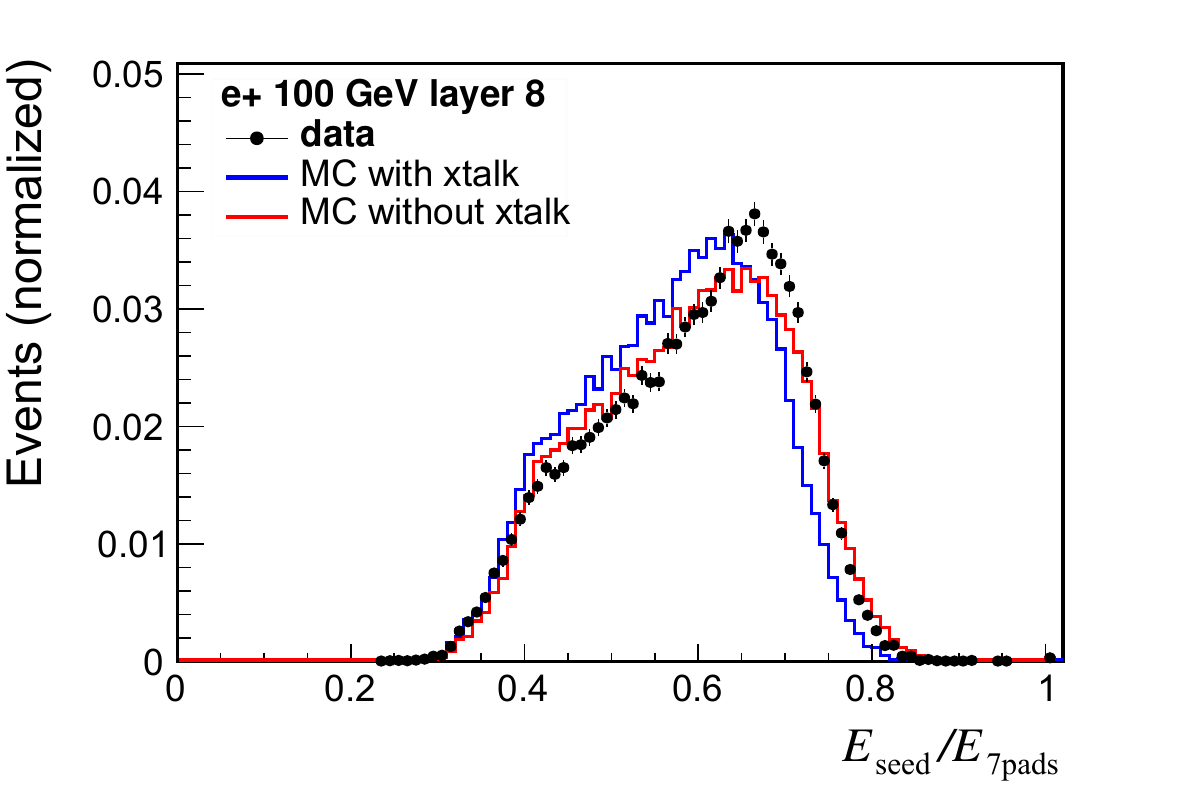}
    \includegraphics[width=0.3\textwidth]{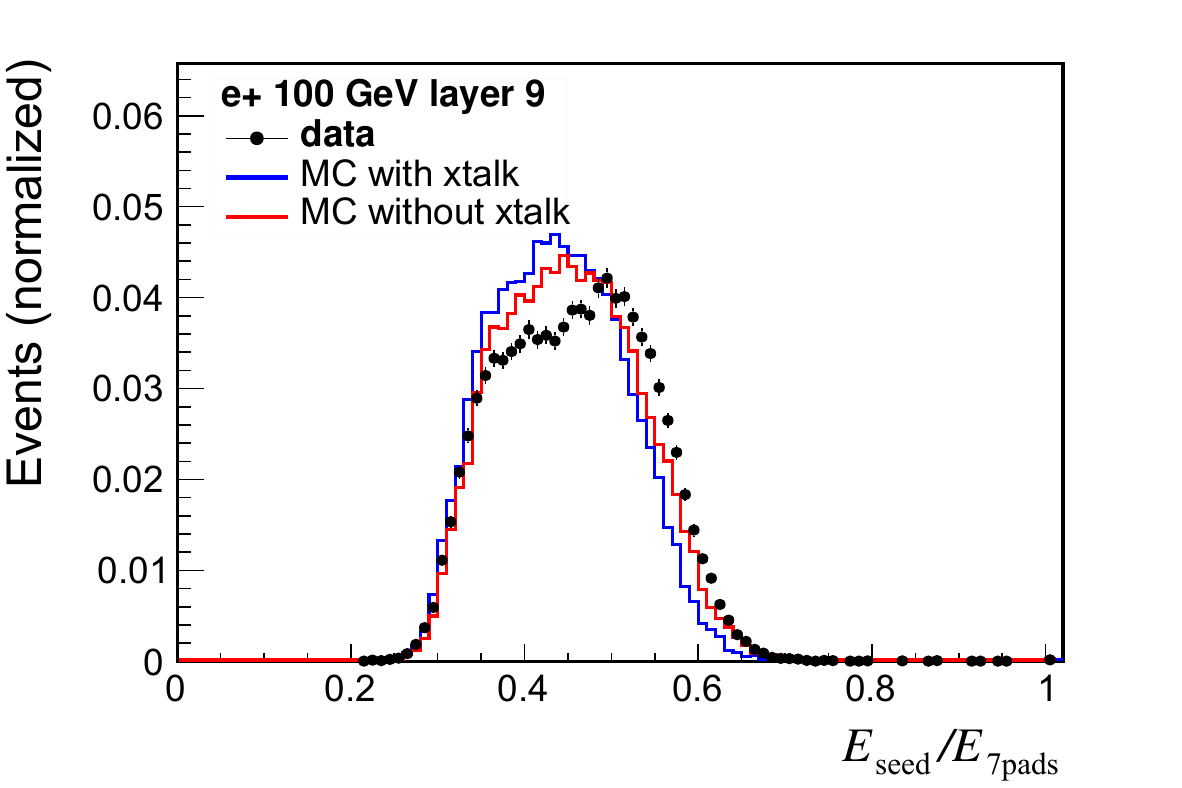}
    \includegraphics[width=0.3\textwidth]{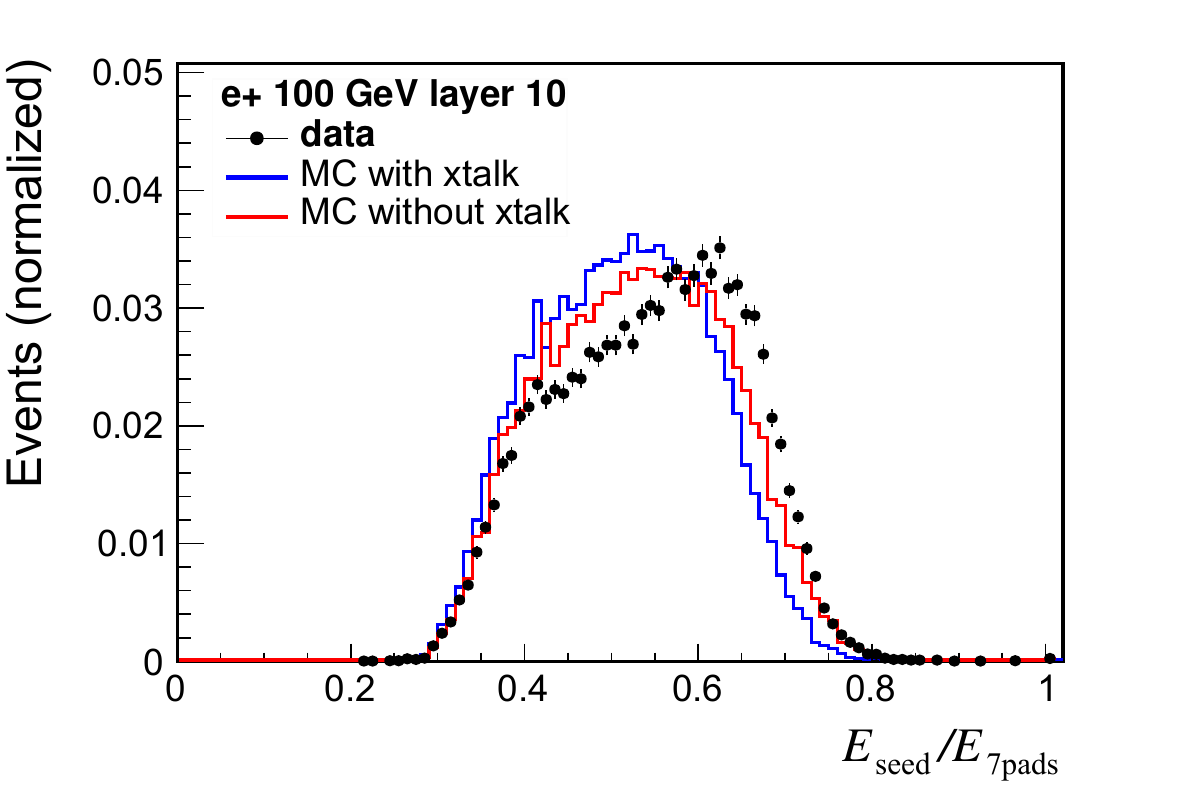}
    \caption{Distributions of $E_{\text{seed}}/E_{\text{7pads}}$ for nominal positron energy of 100~\GeV 
    for layers 3 to 5 and 8 to 10.
    Application of the measured crosstalk in the simulation improves the agreement with the data up to layer 5.}
    \label{fig:e1oe7_xtalk}
  \end{center}
\end{figure*}

The last effect that might influence the lateral shapes is crosstalk. This was studied with a dedicated simulation described in Section~\ref{sec:datasimu}. 
The distributions of $E_{\text{seed}}/E_{\text{7pads}}$ for data and simulation, including or excluding the measured crosstalk, are show in Fig.~\ref{fig:e1oe7_xtalk} for different layers. Inclusion of crosstalk in the simulation led to a better agreement with the data in first layers, going up to the fifth to tenth layer depending on the positron energy. The impact of crosstalk on $E_{\text{7pads}}/E_{\text{19pads}}$ was found to be negligible. 

This improvement was observed independently of the DWC fiducial cuts.
Odd layers with more energy and larger transverse shower shape are less sensitive to crosstalk.
Similarly, the impact of the crosstalk is smaller for layers around the shower maximum. 
In conclusion, even a few-percent crosstalk effect can lead to significant distortions of the lateral shapes.

\begin{figure}[hbtp]
\begin{center}
    \includegraphics[width=0.45\textwidth]{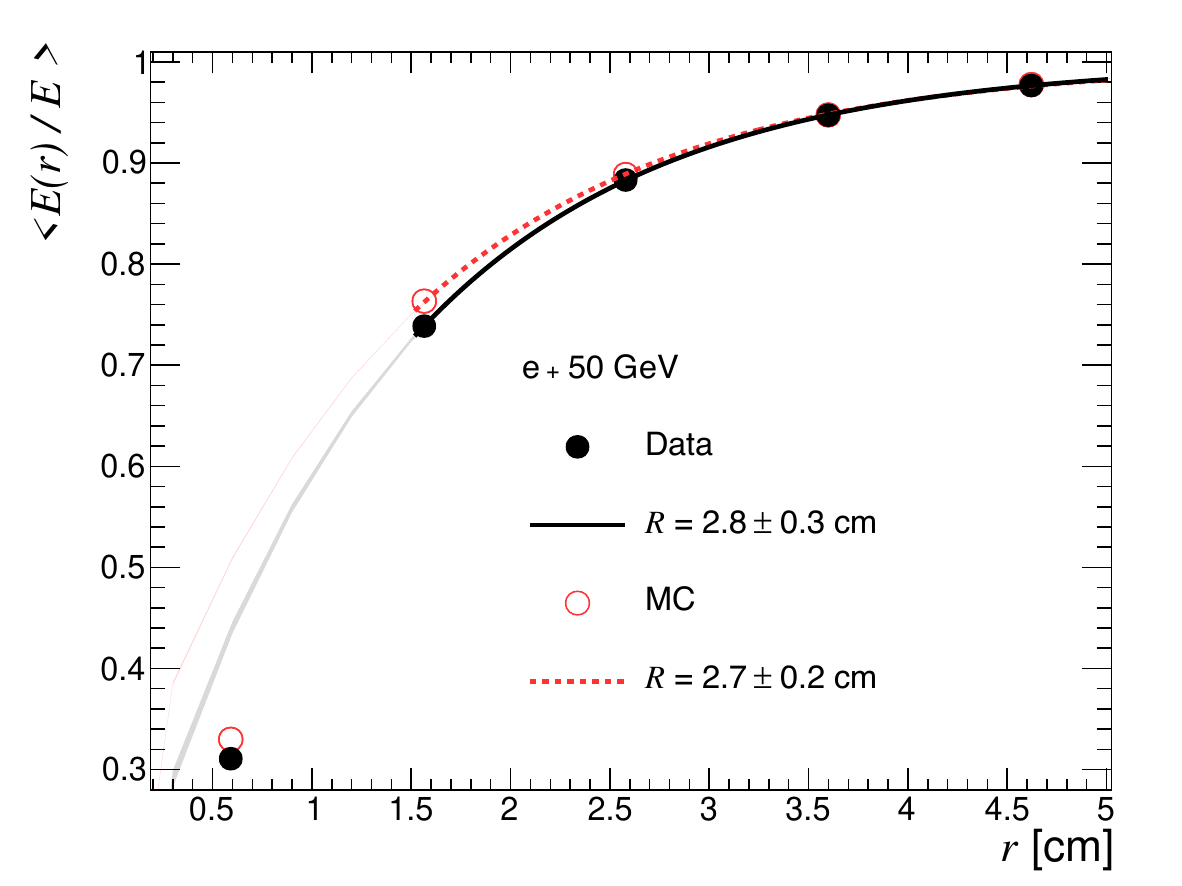}
    \includegraphics[width=0.45\textwidth]{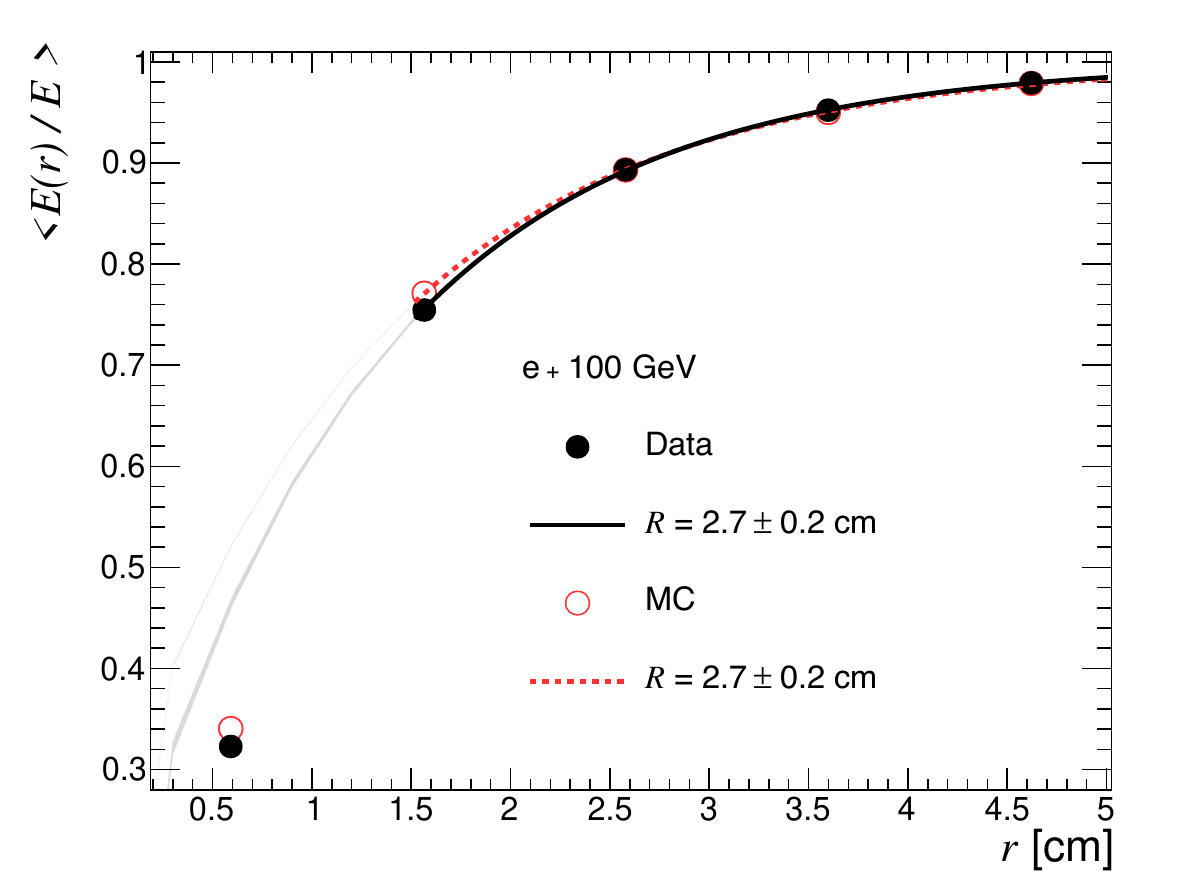}
\end{center}%
\caption{Statistical mean of $E(r)/E$ as a function of $r$ for nominal positron energy of 50 GeV and 100 GeV in data and simulation. The radial containment $R$ is extracted from the fitted exponential function defined in Eq.~\ref{eq:Exponential_fit} using $\left\langle E(R)/E \right\rangle = 0.9$ (the error on $R$ is evaluated by propagating the parameter fit uncertainties and by considering different choices of discrete $r$).}
\label{fig:FractionalEnergy_Full}
\end{figure}

In the environment of the HL-LHC, in which the HGCAL will operate, the showers of single electrons or photons will have to be reconstructed within a significant, uniformly-spread background from pileup collisions. 
Therefore, a small transverse size for electromagnetic showers is desirable for good energy measurement and two-shower separation, particularly at high pseudo-rapidities.
The radial containment of the energy deposition, $R$, was defined as the radius of a cylinder aligned along the shower axis that contains on average 90\% of the energy deposited in the shower.
The energy deposited in a cylinder of radius $r$ aligned along the shower axis, $E(r)$, was evaluated in discrete steps of $r$:
\begin{equation}
E(r) = \sum_{i=1}^{28} E^{\text{Si}}_i(r)
,
\end{equation}
where $E^{\text{Si}}_i(r)$ is, for the $i^{\text{th}}$ layer, the sum of the energy deposited in Si pads inside a given ring of pads centered on the shower axis. 
The corresponding discrete steps of $r$ are such that the area of the disc of radius $r$ corresponds to the total area of the pads used to compute $E^{\text{Si}}_i(r)$.
Here the shower axis is the track trajectory evaluated from the DWCs for each event.
Figure~\ref{fig:FractionalEnergy_Full} shows, for two positron energies, the statistical mean of $E(r)/E$ over all events, where $E$ is the total measured energy of the calorimeter. It can be seen that 90\% of the energy is contained in a cylinder with a radius of about 3~\cm, which corresponds to the central pad surrounded by two rings of 1.1 cm$^ 2$ pads.
In order to obtain $\left\langle E(r)/E \right\rangle$ for all values of $r$, this quantity was parameterized with the following function:
\begin{equation}
\left\langle E(r)/E \right\rangle = 1 - A \cdot \exp(-B \cdot r),
\label{eq:Exponential_fit}
\end{equation}
where $A$ and $B$ are free parameters. 
Having obtained $A$ and $B$ from the fit to $\left\langle E(r)/E \right\rangle$ as a function of $r$, the value of $R$ was obtained by solving $\left\langle E(R)/E \right\rangle$ = 0.9.
The first points in Fig.~\ref{fig:FractionalEnergy_Full}
were not included in the fits as the variation in the deposited energy in the central cell is a strong function of the impact point. The contribution to the error on $R$ from the choice for the discrete $r$ values is significantly larger than the contribution from the fit uncertainties.
For positron energies larger than 50~\GeV, a good agreement between data and MC was observed.
For energies between 20~\GeV and 50~\GeV, the value of $R$ was found to increase with decreasing positron energy for both data and simulation, which is understood to be mainly due to the increase of scattering in the beamline upstream of the calorimeter \cite{CMS_NOTE}. In the same energy range, the simulation predicted a smaller value for $R$, which could be due to an incomplete description of the beamline in the simulation, resulting in less upstream scattering. At high energy, the radial containment is slightly higher than the computed Moli\`{e}re radius of the HGCAL design \cite{CMS_HGCAL_TDR}. 
This difference can be understood to be the result of the larger air gaps between the sampling layers in the prototype than in the HGCAL design.

\section{Conclusion}
\label{sec:conclusion}
%
%
The performance of a 28-layer electromagnetic HGCAL prototype with a lateral segmentation of 1.1$\cm^2$ was studied in a beam test at CERN. The detector was exposed to positron beams with nominal energies ranging from 20 to 300~GeV. 

Direct comparison between data and simulation of the measured energies shows that the data 
is less than the simulation by 3.5\%, independent of the beam energy.
Correcting for this overall scale, the agreement between data and simulation was found to be at the level of 2\% for the full energy range. 
The stochastic term in the measured energy resolution is 0.22~$\sqrt{\text{GeV}}$ and the constant term is 0.6\%, in close agreement with simulation.
The linearity of the energy response is better than 2.5\%. 
Similar performance was obtained using the reconstructed energy deposited in the CE-E prototype estimated with two alternative methods.

The spatial and angular resolutions were also studied as a function of the beam energy.
At the highest energy, using the reconstructed shower axis, the lateral position resolution was estimated to be less than 0.3~mm, and the angular resolution was found to be in the order of 4.5~mrad.

Lateral and longitudinal shower shapes were  measured and compared to the simulation. Good agreement was found in the case of longitudinal shower shapes for all energies, 
with the 3.5\% overall difference between data and simulation located in the core of the shower.
The empirical parameterization matched well the longitudinal shower shapes. For the first layers, the lateral shower shapes show a sensitivity to pad-to-pad crosstalk, which, when included in the simulation, improves the agreement with the data. 

\acknowledgments
We thank the technical and administrative staffs at CERN and at other CMS institutes for their contributions to the success of the CMS upgrade program. We acknowledge the enduring support provided by the following funding agencies and laboratories: BMBWF and FWF (Austria); CERN; CAS, MoST, and NSFC (China); MSES and CSF (Croatia); CEA, CNRS/IN2P3 and P2IO LabEx (ANR-10-LABX-0038) (France); SRNSF (Georgia); BMBF, DFG, and HGF (Germany); GSRT (Greece); DAE and DST (India);  MES (Latvia); MOE and UM (Malaysia); MOS (Montenegro); PAEC (Pakistan); FCT (Portugal); JINR (Dubna); MON, RosAtom, RAS, RFBR, and NRC KI (Russia); MoST (Taipei); ThEP Center, IPST, STAR, and NSTDA (Thailand); TUBITAK and TENMAK (Turkey); STFC (United Kingdom); and DOE (USA).


\bibliographystyle{unsrtnat}
\bibliography{main_jinst}



\end{document}